\documentclass[aps,prb,preprint,onecolumn,superscriptaddress]{revtex4-2}
\usepackage{amssymb,amsmath,amsfonts}
\usepackage{booktabs}
\usepackage{graphicx} 
\usepackage{float} 
\usepackage{natbib} 
\usepackage{color,soul}
\usepackage{xcolor}
\usepackage[colorlinks=true,bookmarks=false,citecolor=blue,urlcolor=blue]{hyperref}
\usepackage{array}
\usepackage{lmodern} 
\usepackage{soul}

\graphicspath{ {./figures/} }

\newcommand{\br}{\mathbf{r} }

\newcommand{\w}{\omega }

\newcommand{\tot}{ \mathrm{tot} }

\newcommand{\rad}{ \mathrm{rad} }

\begin{document}

\title{Deeply Subwavelength Blue-Range Nanolaser}

\newcommand{\HEU}{
Qingdao Innovation and Development Center, Harbin Engineering University, Qingdao 266000 Shandong, China
}

\newcommand{\moscow}{
Moscow Center for Advanced Studies, Moscow 123592, Russia
}

\newcommand{\itmo}{School of Physics and Engineering, ITMO University, St. Petersburg 197101, Russia}

\author{Daria Khemelevskaia}
\affiliation{\itmo}

\author{Nikolai Solodovchenko}
\affiliation{\itmo}

\author{Elizaveta Sapozhnikova}
\affiliation{\itmo}

\author{Igor Chestnov}
\affiliation{\itmo}

\author{Alexey Dmitriev}
\affiliation{\itmo}

\author{Vanik Shahnazaryan}
\affiliation{\moscow}

\author{Denis Baranov}
\affiliation{\moscow}

\author{Sergey Makarov}
\email{s.makarov@metalab.ifmo.ru}
\affiliation{\itmo}
\affiliation{\HEU}

\begin{abstract}
Modern high-definition display and augmented reality technologies require the development of ultracompact micro- and nano-pixels with colors covering the full gamut and high brightness. In this regard, lasing nano-pixels emitting light in the spectral range 400-700~nm are highly demanded. Despite progress in red, green, and ultraviolet nanolasers, the demonstrated blue-range (400-500~nm) single-particle-based lasers are still not subwavelength yet. Here we fabricate CsPbCl$_3$ cubic-shaped single-crystal nanolasers on a silver substrate by wet chemistry synthesis, producing their size range around 100-500~nm, where the nanoparticle with sizes 0.145$\mu$m$\times$0.195$\mu$m$\times$0.19$\mu$m and volume 0.005~$\mu$m$^3$ (i.e. $\sim\lambda^3$/13) is the smallest nanolaser among the lasers operating in the blue range reported so far, with emission wavelength around $\lambda\approx 415$~nm. Experimental results at a temperature of 80~K and theoretical modeling show that the CsPbCl$_3$ nanolaser is a polaritonic laser where exciton-polaritons are strongly coupled with Mie resonances enhanced by the metallic substrate. As a result, the combination of the strong excitonic response of CsPbCl$_3$ materials, its high crystalline quality, and optimized optical resonant properties resulting in a population-inversion-free lasing regime are the key factors making the proposed nanolaser design superior among previously reported ones in the blue spectral range.
\end{abstract}

\maketitle
\newpage

\section{Introduction}
Blue nanolasers operating in the blue part of the visible spectral range ($\lambda =400-500$~nm) are promising types of light emitters for various applications in high-definition displays, high-density optical storage, biomedical imaging, and quantum communication~\cite{ma2019applications, azzam2020ten}. 
However, pushing small lasers to nanoscale sizes is challenging due to the diffraction limit, which for photonic cavities typically sets a minimum volume on the order of $(\lambda/2n)^3$ (where $n$ is the refractive index), while plasmonic-based cavities highly localizing modes suffer from additional losses. The near-UV nanolasers ($\lambda = 350-400$~nm) and green ($\lambda =500-550$~nm) were successfully demonstrated basing on various wide-bandgap semiconductors as gain media, where such key material systems were employed: III-nitrides (GaN, InGaN)~\cite{gradevcak2005gan, zhang2014room} and wide-gap II–VI semiconductors (ZnO, CdS, ZnCdS, ZnSe)~\cite{huang2001room,duan2003single,oulton2009plasmon, zhuang2012composition}. However, there were no demonstrations of truly subwavelength nanolasers (in all three dimensions) for the spectral range $\lambda =400-500$~nm. In turn, organic materials like oligofluorenes~\cite{chang2021tuning} enable low-threshold, air-stable deep-blue lasing but suffer from concentration quenching and cost-effective processing during the formation of a subwavelength nanocavity.


Halide perovskites (ABX$_3$, where A - Cs,MA,FA and X - Cl,Br,I) are solution-processable materials supporting high optical gain~\cite{tatarinov2023high}, low lasing thresholds, and tunable output covering the entire optical range~\cite{shi2025ten}. In particular, the perovskites with mixed anion composition (e.g., CsPb(Br/Cl)$_3$) can support emission in the blue spectral range. Recent advances in halide perovskites providing high PL quantum yields (PLQY) up to 90\% and perovskite-based LEDs achieved external quantum efficiencies up to EQE$\approx$13\% via vapor-assisted crystallization~\cite{wang2020dimension, karlsson2021mixed, liu2021water, sun2022bifunctional}. 

\begin{figure}[h!]
    \centering
    \includegraphics[width=0.6\linewidth]{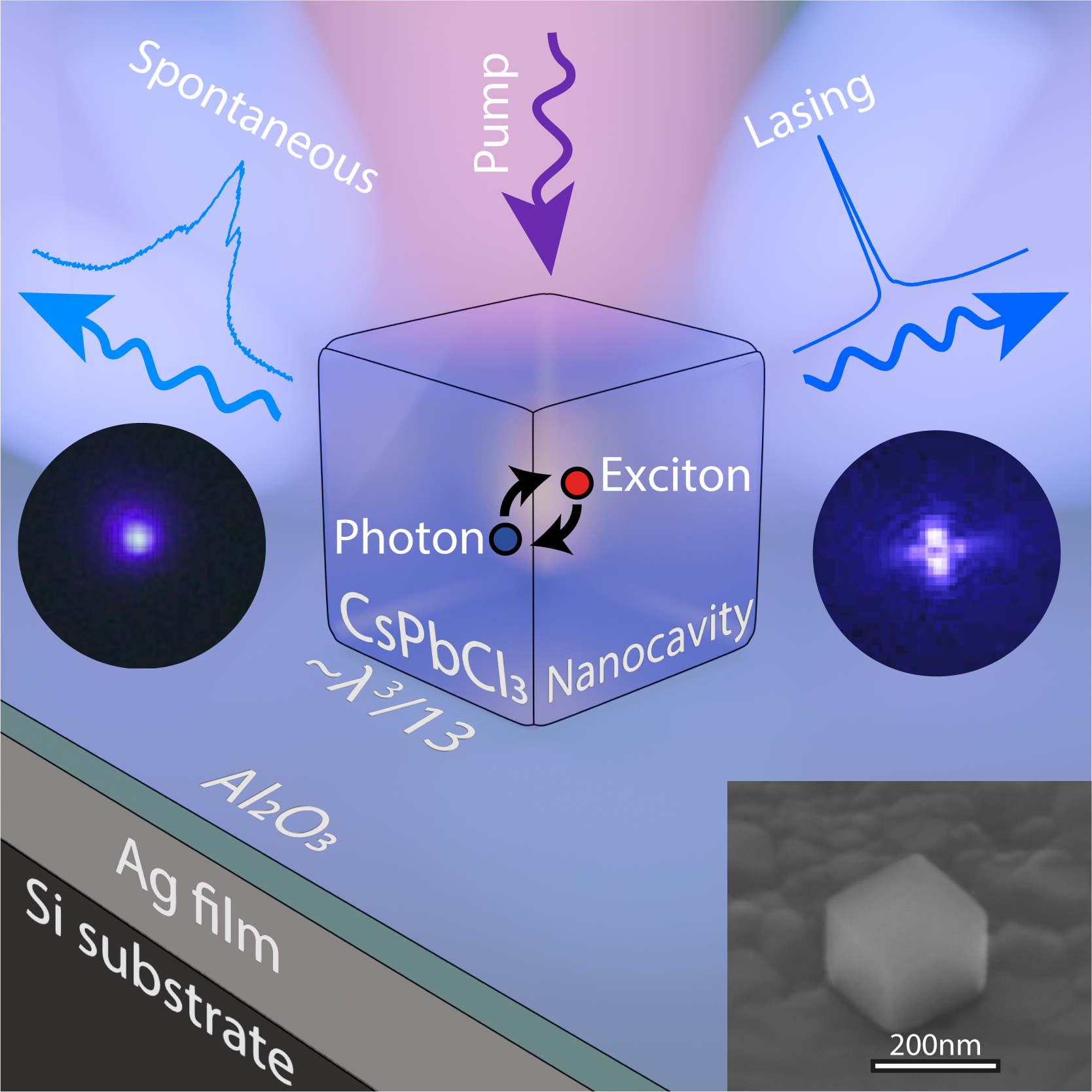}
    \caption{\textbf{Design.} Illustration of the experimental sample representing a perovskite polariton nanolaser on a silver film with Al$_2$O$_3$ spacer. Insets: PL from a perovskite nanolaser below (left inset) and above (right inset) lasing threshold with characteristic spectra shown above the arrows; SEM image shows a typical perovskite CsPbCl$_3$ nanolaser on the silver film.}
    \label{fig:0}
\end{figure}

In this work, we develop the smallest nanolaser reported so far for the blue spectral range (i.e. $\lambda\approx 400-500$~nm). The nanolasers are made of CsPbCl$_3$ halide perovskite nanocubiods fabricated from a chemical reaction in solution and deposited on a silver film (see the design in Figure~\ref{fig:0}). The ratio between the lasing wavelength ($\lambda\approx 415$~nm) and volume ($V \approx 0.005 \mu$m$^3$) makes it deeply subwavelegth and, moreover, smaller than all previously reported perovskite-based nanolasers with various compositions. According to our experimental results and theoretical calculations, the mechanism of the developed nanolaser operation is the condensation of exciton polaritons at the Mie resonance. Such a compact laser exhibits a relatively low optical threshold (around $10 \mu$Jcm$^{-2}$) at a temperature of 80~K, which makes its application in various photonic-chip tasks feasible.

\label{ch:results}
\section{Results}

\subsection{Temperature-dependent photoluminescent properties} 

\begin{figure}[t!]
    \centering
    \includegraphics[width=0.8\linewidth]{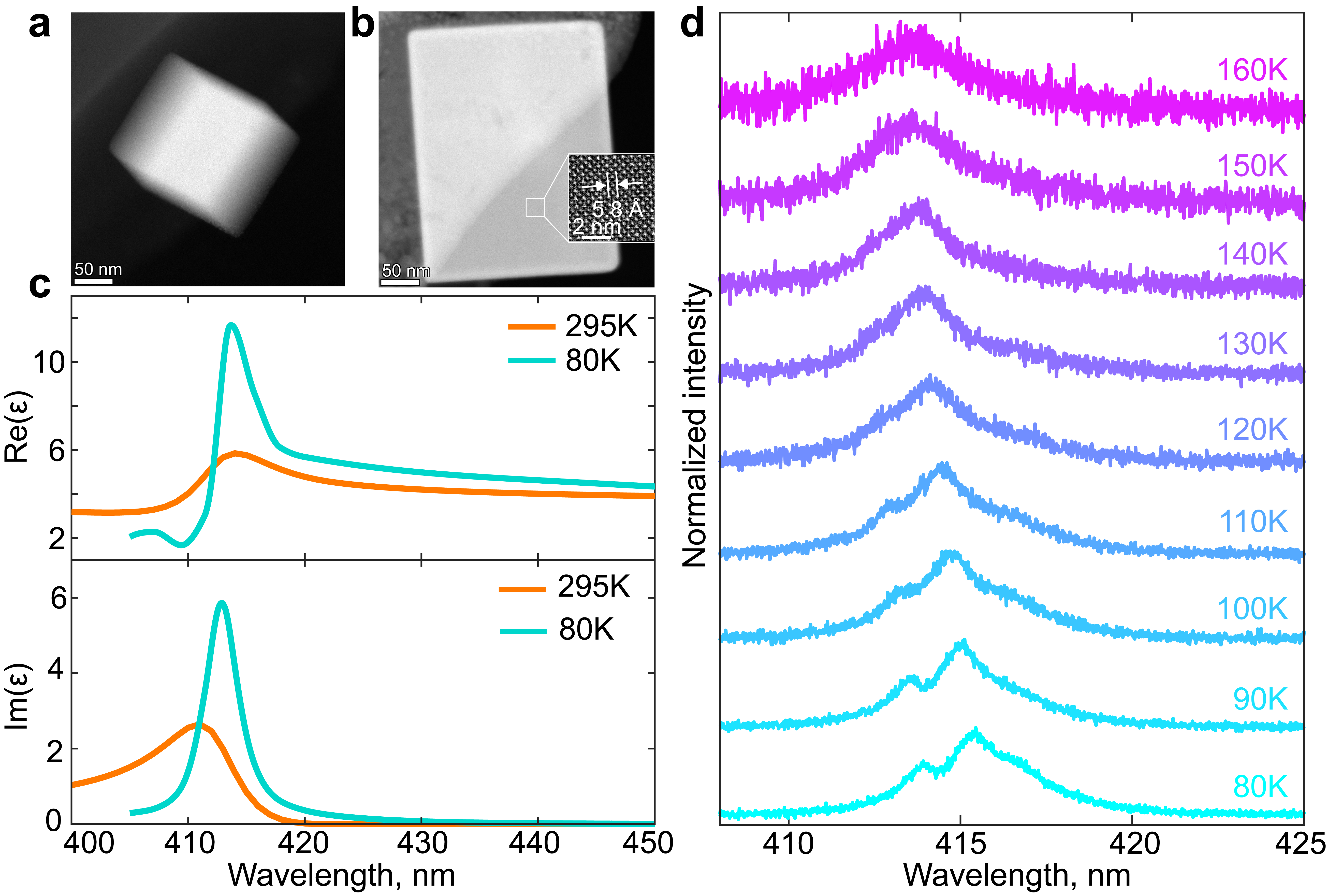}
    \caption{\textbf{Optical properties of CsPbCl$_3$ nanocubes.} (\textbf{a,b}) STEM HAADF images of nanocubes. (\textbf{c}) The real and imaginary parts of dielectric permittivity at room (reconstructed from the work~\cite{ermolaev2023giant}) and nitrogen temperatures . (\textbf{d}) Normalized PL spectra obtained at different temperatures exhibiting gradual transition from a single peak at $160$~K to two separate maxima upon cooling down to $80$~K.}
    \label{fig:1}
\end{figure}

Monohalide CsPbCl$_3$ nanocubes were prepared according to a modified hot-injection approach (see Methods) and were then integrated with a metal-dielectric substrate Al$_2$O$_3$/Ag/Si. The obtained nanocubes have regular shapes and crystalline morphology, as confirmed by Scanning Transmission Electron Microscopy with High-Angle Annular Dark-field (STEM HAADF), as shown in Figure~\ref{fig:1}a,b. Moreover, the elemental composition of CsPbCl$_3$ nanocubes was confirmed by means of energy dispersive X-ray spectroscopy (Fig.S3). 

It is well established that CsPbCl$_3$ perovskite exhibits bright photoluminescence (PL) with a Gaussian profile around $\lambda\approx 413$~nm at room temperature, while the exciton binding energy can be as high as $65$~meV~\cite{baranowski2020exciton}. Recently, a strong enhancement of excitonic properties was revealed for CsPbBr$_3$ at low temperatures~\cite{masharin2024polariton} taking into account room-temperature measurements published elsewhere~\cite{ermolaev2023giant}. Similarly to the previous works, we estimated a wavelength-dependent dielectric permittivity for CsPbCl$_3$ at temperature T=80~K (see Fig.S2 for more details). A significant increase in permittivity was obtained near the exciton resonance compared to the value at room temperature (Figure~\ref{fig:1}c), which is caused by an increase in the amplitude strength of the excitonic resonance.

To study the effect of temperature on the PL properties of the samples, they were placed in a nitrogen cryostat chamber. Femtosecond (fs) laser pulses with a wavelength of $\lambda = 395$~nm and a repetition rate of $\mathrm{RR} = 100$~kHz were used as the excitation source (see Methods for details on the experimental setup). Temperature-dependent measurements were conducted at a fixed excitation fluence of 1.5~$\mu$J$\cdot$cm$^{-2}$ in the reflection configuration (see Fig. S7 for experimental details). The emission signal collected from a single perovskite nanocube maintained a conventional shape when cooled down to $T=150$~K (Figure~\ref{fig:1}d). Interestingly, when the temperature was reduced below $T=140$~K, the PL profile became unevenly broadened, whereas several separate peaks became distinguishable at lower temperatures ($T=80-90$~K). In order to explain this unusual effect, we further study modulated PL spectra at $T=80$~K from nanocubes of different sizes and performed a theoretical description by means of a macroscopic model.

\subsection{Macroscopic model of photoluminescence spectra} 

\begin{figure}[t!]
    \centering
    \includegraphics[width=1\linewidth]{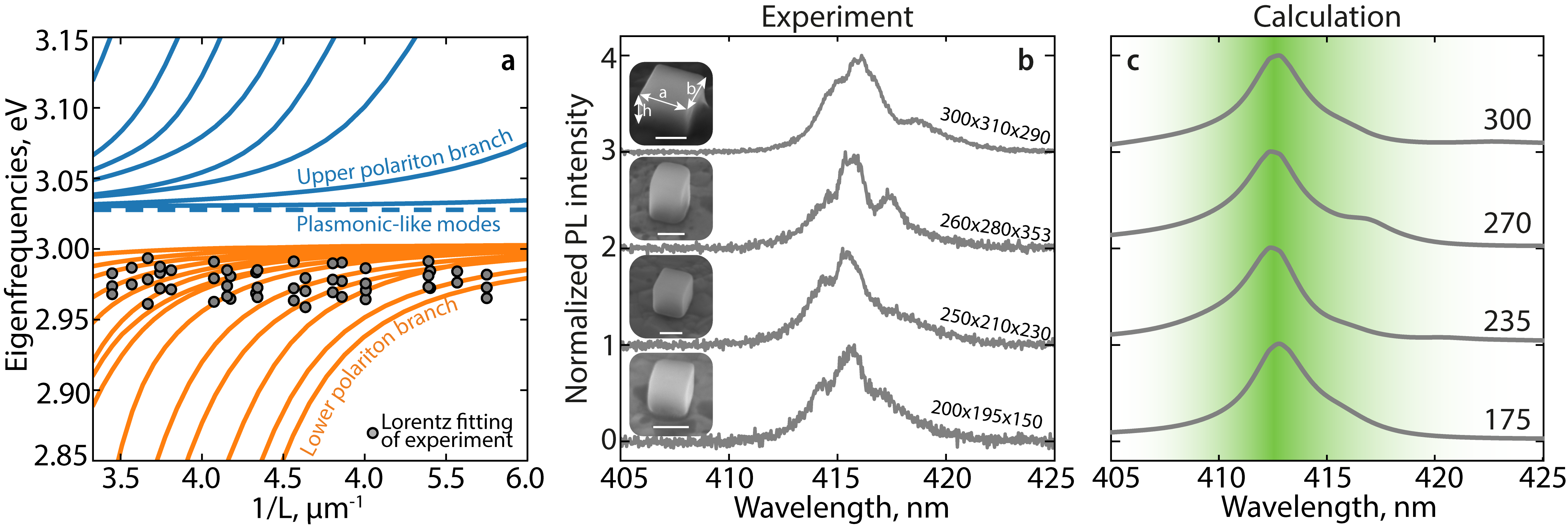}
    \caption{\textbf{PL splitting in CsPbCl$_3$ nanocubes.} 
    (\textbf{a}) Calculated polariton branches based on QNM frequencies (solid lines) and maxima of Lorentz peaks based on fitting of experimental PL spectra obtained at $80$~K (dots). The blue dotted line indicates plasmon-like modes. 
    (\textbf{b}) Experimental PL spectra for nanocuboids with different sizes. The insets show SEM images of the nanocuboids, where the white scale bar corresponds to 200~nm. Geometrical dimensions ($a$ $\times$ $b$ $\times$ $h$) are indicated on the right above each spectrum in nanometers. 
    (\textbf{c}) Calculated spectra for nanocubes with averaged geometry from panel (b). Green gradient schematically illustrates the bare exciton emission spectrum $S(\w).$}
    \label{fig:2}
\end{figure}

Numerical simulation of PL spectra of perovskite nanocubes in the low-intensity pump regime was performed with the use of the quasinormal modes (QNM) theory~\cite{Sauvan2013,lalanne2018light}. The search for the QNMs of the perovskite nanocube is based on a recently developed auxiliary field formalism~\cite{yan2018rigorous, wu2020intrinsic}, which works effectively for resonators containing dispersive media. Searching for QNMs of the nanocube with use of the dispersive permittivity of CsPbCl$_3$ and the substrate (see Methods) yields the eigenfrequencies of the perovskite nanocubes as a function of the cube size $L$ (Figure~\ref{fig:2}a). The spectrum consists of a rich set of upper (blue curves) and lower (orange curves) polariton branches, exhibiting an avoided crossing behavior and separated by the energy gap being typical for exciton-polariton dispersions. 
This multitude of polaritonic branches is the result of the interaction of the CsPbCl$_3$ exciton at $2.995$~eV with the bare photonic modes of the weakly dispersive nanocube having only the high-energy Lorentz pole switched on (Fig. S5a). Additionally, the dashed line shows the dispersion of plasmon-like modes of the cube, associated with negative-permittivity region of the Lorentz model in the complex frequency plane (Fig. S3).
This picture contrasts spectra of more traditional two-component polaritonic systems with a separate external cavity, which typically show a well-resolved single pair of upper and lower polaritons~\cite{Yoshle2004, Reithmaier2004, Gross2018}.

Circles in Fig.~\ref{fig:2}a denote the frequencies extracted by fitting the experimental PL spectra with Fano contours for nanocubes of the corresponding size; the number of points for each $L$ corresponds to the number of Lorentzian contours used in fitting the PL. As one can see, all frequencies in the experimental PL spectra occur in the region of lower polariton modes; no PL maxima were observed in the region of upper polariton modes above $3.027$~eV. 
This is in line with previous observations of PL splitting in polaritonic systems, which also demonstrate prominent PL from lower polariton modes, and at the same time strongly suppressed PL from upper polariton modes~\cite{Bellessa2004, Wersall2017, Wersaell2019, shang2018surface}.

Having obtained the QNM spectra of the perovskite nanocubes, we move on to the evaluation of their PL spectra.
These QNMs obtained by solving macroscopic Maxwell's equations are essentially collective in nature: they are the result of the interaction of the volume distribution of the exciton dipole moment with the photonic modes of the nanocube, as in self-hybridized polaritonic systems~\cite{Canales2021, Weber2023}.
Furthermore, each single exciton is only in the weak coupling regime with its optically resonant environment.
This allows us to model each infinitesimal volume of the nanocube as an elementary point emitter and describe its spontaneous emission through the local density of states LDOS$(\br, \w)$ using the Fermi's golden rule~\cite{Blanco2004, anger2006enhancement, Pelton2015}. This is in contrast to single-emitter polaritons, where a single electronic transition of a molecule or a quantum dot interacts non-perturbatively with an optical cavity~\cite{Yoshle2004, Reithmaier2004, Chikkaraddy2016, Gross2018, Park2019, Hu2021}. Finally, the PL signal from the entire cube can be obtained by integrating the radiative decay rate enhancement factor $F_\rad(\br, \w)$ over the entire volume of the cube with weighting factors~\cite{ringler2008shaping}:
\begin{equation}
    \mathrm{PL}(\w, \w_{pump}) = 
    \int_{V} \mathrm{EF}(\br,\w_{pump}) 
    \frac{ S(\w) F_\rad (\br, \w)}{ \Gamma (\br) } d^3 \br,
    \label{eq:PL_integral}
\end{equation}
where $\mathrm{EF}(\br,\w_{pump}) = \left| \mathbf{E}/\mathbf{E}_0 \right|^2$ is the pump field enhancement at the laser frequency at each point in the cube volume (Fig. S4), $\mathbf{E}_0$ is the electric field of the pump field in the absence of the perovskite resonator, $\Gamma (\br) = \int_0 ^\infty S (\w) F_\tot (\br, \w) d\w$ is the position-dependent total decay rate enhancement factor, and $S(\w)$ is the spectral envelope of the bare exciton emission into non-resonant photonic environment:
\begin{equation}
    S(\w) = \frac{\gamma_{ex}^2}{(\w-\w_{ex})^2 + \gamma_{ex}^2},
\label{Exiton_lotentz}
\end{equation}
where $\w_{ex}$ and $\gamma_{ex}$ represent the resonant frequency and linewidth of the low-energy Lorentz pole responsible for the exciton emission (see Methods). Figure~\ref{fig:2}b shows the calculated PL intensity map as a function of the cube size $L$. The white dotted lines indicate the lower polariton branches, and the thickness of the lines is proportional to the contribution of each particular QNM to the PL spectrum. 

Figures~\ref{fig:2}c,d show a series of normalized experimental and numerical PL spectra for various cube sizes. The experimental spectra were obtained at a temperature of $T=80$~K and a laser pump fluence of $0.1$~$\mu$Jcm$^{-2}$ at a wavelength of $\lambda=395$~nm. To compare the calculations with the experimental spectra from non-cubic particles, the sizes of the measured cubes were estimated as the average value over three sides $L \approx (a + b + h)/3$. The calculated spectra are in qualitative agreement with the measured PL spectra.

The spectra are conventionally composed of two parts: a long-wavelength peak that disperses with the cube size, and a short-wavelength peak that nearly coincides with the energy of the bare exciton $\w_{ex}$. The second part of emission originates from the high density of closely packed lower polariton states just below the exciton energy (Fig.~\ref{fig:2}a), which all merge into a single emission line representing the uncoupled exciton~\cite{Bellessa2004, Canales2021}.

Using the calculated single-pole quasinormal modes (Fig. S5a), we also calculated artificial PL spectra from nanocubes of various sizes as if the emission was originating from a weak oscillator strength transition, whose presence would not modify the permittivity dispersion and the eigenfrequencies spectrum.
The resulting artificial PL spectra do not demonstrate any kind of polaritonic features (Fig. S5b) corroborating the conclusion that the observed splitting in the PL spectra (Fig.~\ref{fig:2}b) can indeed be attributed to the formation of polaritonic modes in a CsPbCl$_3$ nanocuboid cavity.

\subsection{Polariton lasing in perovskite nanocubiods} 

\begin{figure}[h!]
    \centering
    \includegraphics[width=1\linewidth]{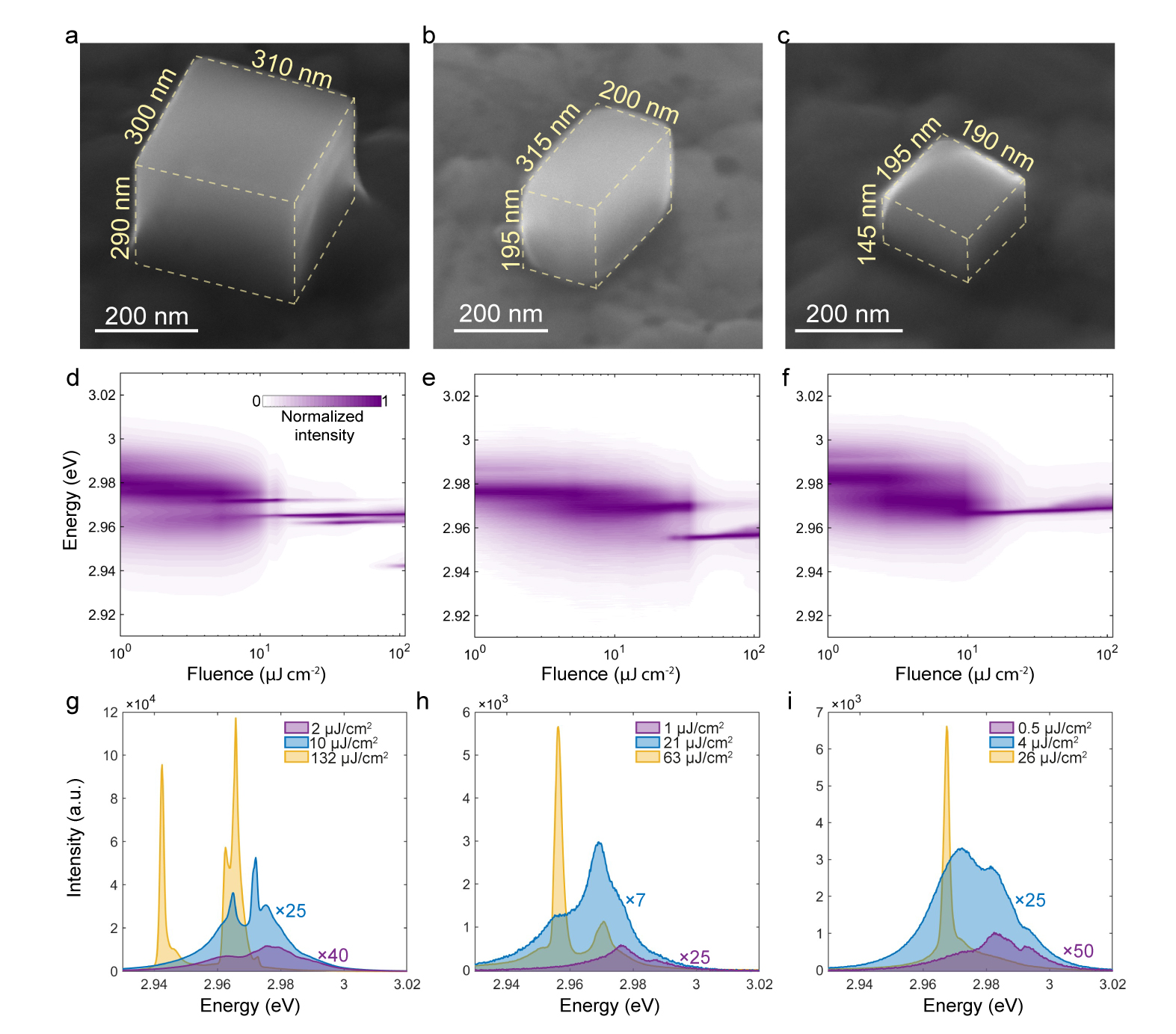}
    \caption{\textbf{Polariton lasing at $80$~K.} (\textbf{a-c}) SEM images of studied perovskite nanocuboids with different sizes. Their physical volume are 0.031, 0.012, and 0.005 $\mu$m$^3$, respectively. (\textbf{d-f}) Normalized emission spectra depending on the pump fluence obtained at $80$~K for corresponding nanocubes. (\textbf{g-i}). The emission spectra obtained at low, medium, and high incident fluences indicated by the corresponding colors.}
    \label{fig:3}
\end{figure}

We further study the behavior of nonuniform PL spectra of CsPbCl$_3$ nanocubes at elevated pump fluence. For this, three nanocubes with different sizes, depicted in Figure~\ref{fig:3}a-c, were excited by fs-laser pulses with $\lambda = 395$~nm at $T=80$~K (see Methods and Fig.S3 with experimental setup). According to SEM, the biggest cube is around 300 by 310 by 290 nm (Figure~\ref{fig:3}a), the second one is $315\times200\times195$~nm (Figure~\ref{fig:3}b), and the smallest one is only $195\times190\times145$~nm with V~$\approx0.005$~$\mu$m$^3$ (Figure~\ref{fig:3}c).

Pump-dependent PL spectra measured for corresponding cubes are presented in Figure~\ref{fig:3}d-f. The PL intensity of each fluence was normalized by its maximum for clarity. It clearly shows that the maximum PL intensity begins to redshift at higher pump fluences. As was discussed above, the PL spectra of nanocubes represent the superposition of closely packed lower polariton states. With increasing optical pumping, the number of polaritons increases, leading to their phonon-mediated scattering downwards in energy and, consequently, to a spectral redistribution of the collected emission. Moreover, at higher pump fluence, the spectra narrow sharply, while almost all emission intensity comes from separate peaks. This phenomenon can be attributed to the transition from spontaneous polariton emission to the polariton lasing regime. Indeed, multimode lasing was obtained for the biggest perovskite cube with excitation above $10$~$\mu$J~cm$^{-2}$. Corresponding emission spectra at different pump fluence is shown in Figure \ref{fig:3}g, demonstrating the reshaping of PL profile and appearance of multiple narrow peaks corresponding to specific low-energy polariton states (Figure~\ref{fig:2}a). The medium nanocube demonstrates similar behavior, but with several modes dominating and producing lasing at pump fluence above $25$~$\mu$J~cm$^{-2}$ (Figure~\ref{fig:3}e,h). Additionally, optical images of nanocubes radiation taken below and above threshold reveals the appearance of interference fringes, indicating the coherence of the polariton lasing emission (Fig.~S5).

\begin{figure}[t!]
    \centering
    \includegraphics[width=0.75\linewidth]{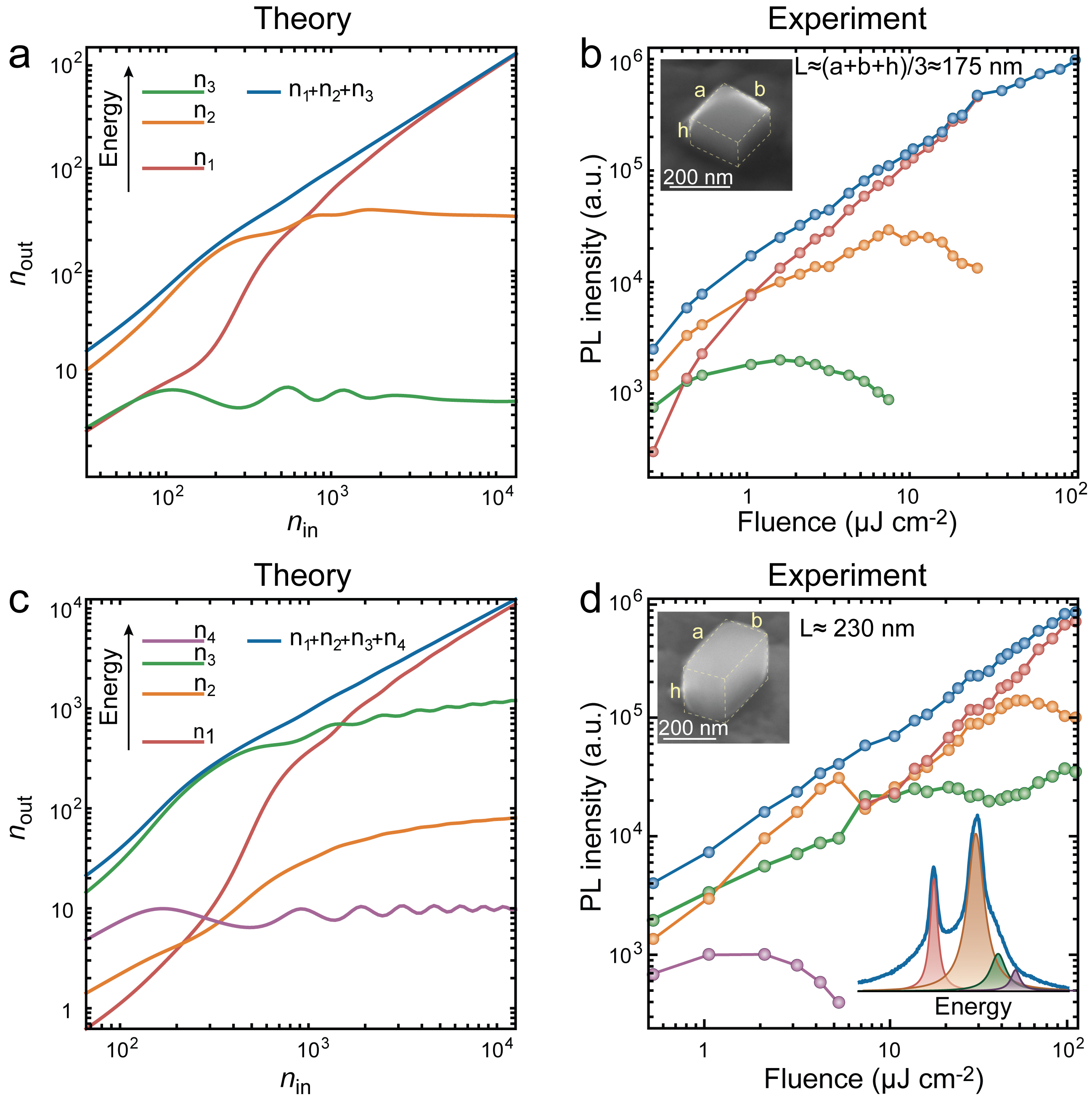}
    \caption{\textbf{Modeling of polariton lasing at $80$~K.} 
    \textbf{(a)} The calculated emission power $n_i^{\rm out} = \tau_i^{-1}\int n_i(t) {\rm d}t$ from the discrete polariton states versus the injected exciton number governed by the pump fluence for the nanocuboid with average size $L=175$~nm. Polariton levels $n_i$ are counted from the lowest in energy, as shown in the label chart.
    \textbf{(b)} Weights of the emission lines extracted from the measured PL spectra of the nanocuboid with the average size $L=175$~nm.
    \textbf{(c)} The same as in panel (a) but for the nanocuboid with average size $L=230$~nm.
    \textbf{(d)} Weights of the emission peaks extracted from the measured spectra of the nanocuboid with average size $L=230$~nm.  Inset: a sketch of the emission spectrum fitted with Lorentzian lines.}
    \label{fig:5}
\end{figure}

On the contrary, the smallest nanocube (Figure~\ref{fig:3}c) reveals single mode lasing at 2.967 eV upon excitation slightly above $10$~$\mu$J~cm$^{-2}$ (Figure~\ref{fig:3}f,i). 
The linewidth of the lasing mode was as narrow as $1.9$~meV (Q$_{las}$~$\sim$~1560), while the intensity versus incident fluence demonstrates a nonlinear increase (Figure \ref{fig:5}b). According to the eigenmode analysis, laser generation occurs at the polariton mode with a frequency of 2.954 eV and a quality factor of about $Q \approx$~61.5 (Fig.~S6). 

\textbf{Theory of pump-dependent emission} 

We study the emission from discrete energy polariton levels.
The collected spectra demonstrate nonlinear features with pump-dependent redistribution of the emission towards the lowest energy state.
This implies the presence of scattering between different polariton levels.
Given the discrete character of the spectrum with energy gaps of orders of several meVs, we attribute the primary relaxation channel to the exciton scattering on longitudinal optical (LO) phonons.
We model the system dynamics via coupled rate equations on populations of the optically dark exciton reservoir and bright polariton states~\cite{masharin2024polariton}:
\begin{align}
    \label{eq:dynamics}
    \frac{{\rm d} n_{\rm R} }{{\rm d} t}  &=    - \frac{n_{\rm R}}{\tau_{\rm R}}   
    -\sum_i \frac{n_{\rm R}}{\tau_{{\rm R},i} }  (n_i +1)  
    + P_0 e^{-(t-t_0)^2/(2\delta t^2)} , \\
    \frac{{\rm d} n_i }{{\rm d} t} &=  - \frac{n_i}{\tau_i}  
    +\frac{n_{\rm R}}{\tau_{{\rm R},i} }   (n_i +1) 
    + \sum_{j> i} \frac{1}{\tau_{j,i} } \left[  n_j (n_i +1) 
    - e^{-\hbar \omega_{\rm LO} / (k_{\rm B} T)} n_i (n_j +1) \right]  \notag \\
    & +\sum_{j< i} \frac{1}{\tau_{i,j} } \left[  e^{-\hbar \omega_{\rm LO} / (k_{\rm B} T)} n_j (n_i +1) 
    -  n_i (n_j +1) \right]
\end{align}
Here, $n_{\rm R}$ and $\tau_{\rm R}$ denote the total population of reservoir and respective non-radiative lifetime, while $n_i$, $\tau_i$ are the $i$-th polariton state population and lifetime, respectively, with $i =1,2,3,\ldots$ labeling the polariton states in an energy increasing order.
We assume that the radiative decay is the primary channel of polariton relaxation, and discard the non-radiative decay processes. The scattering rates from the reservoir to the bright states $\tau_{{\rm{R}},i}^{-1}$ and between the discrete levels $\tau_{i,j}^{-1}$ are governed by the material parameters (see~\ref{ch:Methods}Methods).
The LO phonon energy $\hbar\omega_{\rm LO} = 26$ meV in CsPbCl$_3$ \cite{liao2019situ}, $k_{\rm B}$ is the Boltzmann constant, $T=80$ K denotes the temperature, and $P_0$ characterizes the pump strength of the pulsed excitation of duration $\delta t$. 
Further details of the approach are presented in~\ref{ch:Methods} Methods.

We consider the emission properties of two nanocuboids having lateral sizes $175$~nm and $230$~nm whose emission spectra reveal the presence of 3 and 4 discrete polariton levels, respectively.
The results of simulations along with the weights of the emission peaks extracted from the experimental spectra are presented in Fig.~\ref{fig:5}.
In Fig.~\ref{fig:5}a the polariton emission versus the injected exciton number is shown for the nanocubooid of lateral size $175$~nm.
There is an accidental resonance of the LO phonon energy and the gap between dark exciton reservoir and the middle polariton level: $E_{\rm X} -E_2^{\rm 175 \, nm} \approx \hbar \omega_{\rm LO}$.
This results in efficient feeding of the middle energy state, $1/\tau_{{\rm R},2}^{\rm 175 \, nm} > 1/\tau_{{\rm R},3}^{\rm 175 \, nm}$. 
As Fig.~\ref{fig:5}a shows, the middle peak dominates in emission spectra at low fluences, as confirmed by the measured spectra shown in Fig.~\ref{fig:5}b. 
The increase of pump fluence triggers stimulated scattering between the discrete bright states, resulting in a gradual enhancement of the population of the lowest energy level, which eventually dominates the emission spectra at the large pump fluence.

The emission of the cube with the edge length $230$~nm reveals a similar behavior. A couple of the highest energy levels detuned from the bare exciton resonance by about $\hbar \omega_{\rm LO}$ dominate at low fluences due to more efficient feeding from the dark states (see Figs.~\ref{fig:5}c and \ref{fig:5}d).
The increase of pump strength stimulates descending scattering towards lower levels, which progressively shifts the emission intensity towards the lowest state, which dominates in the nonlinear regime associated with the onset of lasing. This behavior agrees with the measured spectra shown in Fig.~\ref{fig:5}d.

\section{Conclusion}

Basing on the developed fabrication approach and nanophotonic design optimization, we have created the smallest nanolaser among the devices operating in the blue range ($400-500$~nm) reported so far, with emission wavelength around 415~nm. Namely, the perovskite CsPbCl$_3$ nanolasers with sizes down to a volume of 0.005~$\mu$m$^3$ (or $\lambda^3$/13) have been placed on a silver substrate covered by a dielectric spacer for chemical protection. Experimental results at a temperature of 80~K and theoretical modeling show that the CsPbCl$_3$ nanolaser exhibits a polaritonic nature, where exciton-polaritons are strongly coupled with Mie resonances enhanced by the metallic substrate. We believe that the combination of the strong excitonic response of CsPbCl$_3$ materials, its high crystalline quality, and optimized optical resonant properties resulting in a population-inversion-free lasing regime are the key factors making the proposed nanolaser design superior among previously reported ones in the blue spectral range.


\section{Methods}\label{ch:Methods}

\textbf{Perovskite nanocube synthesis}

Mie-resonance CsPbCl$_3$ nanocuboids were prepared according to the protocol that reproduced a hot injection method~\cite{masharin2024polariton} with slight modifications. Namely, 0.2 mmol of PbCl$_2$ was stirred (1,000 rpm) in 7 ml of diphenyl ether in a vacuum-stimulated two-neck round-bottom flask (50 ml) at 120 $^{\circ}$C for 1 h. Subsequently, the flask was filled with N$_2$ gas, heated to 150 $^{\circ}$C, and 300 $\mu$L (0.91 mmol) of oleylamine (OLAm) and 300 $\mu$L (0.95 mmol) of OA were consequently added dropwise to dissolve lead chloride powder. The obtained solution was heated to 180 $^{\circ}$C followed by rapid injection of 0.5 mL (0.05 mmol) of preheated CsOA in ODE solution. In particular, CsOA was preprepared as described in~\cite{masharin2024polariton} and stored in the glovebox with a nitrogen atmosphere. The reaction mixture was stirred at 180 $^{\circ}$C for 40 min and then quenched in an ice bath. The resultant suspension in diphenyl ether was mixed with n-hexane in a ratio of 1:5. Finally, the solution was centrifuged at 500 rpm for 5 min, and the supernatant was pipetted out, while the sediment with Mie-resonance nanocuboids was redispersed in 8 mL of n-hexane for further utilization.

\textbf{Device fabrication}

Metal dielectric Al$_2$O$_3$/Ag/Si substrates were manufactured according to the method described elsewhere~\cite{masharin2024polariton}. The suspension of CsPbCl$_3$ nanocuboids in n-hexane was dropcast on a metal-dielectric substrate to carry out further optical studies. The morphology and size of the nanocuboids were studied using a Zeiss Merlin scanning electron microscope. HAADF-STEM images were obtained on a Titan Themis Z transmission electron microscope equipped with a DCOR+ condenser spherical aberration corrector at an accelerating voltage of 200 kV. Elemental mapping and X-ray Energy Dispersive spectroscopy (EDX) spectra were collected in STEM mode using a Super-X quad wide-angle X-ray detector array.

\textbf{Optical measurements}

Femtosecond (fs) laser (Pharos, Light Conversion) coupled with a broad-bandwidth optical parametric amplifier (Orpheus-F, Light Conversion) was used as an excitation source ($\lambda_{exc}$ = 395~nm, RR = 100~kHz, $\tau$ = 220~fs). The samples were placed in a nitrogen cryostat (Linkam Scientific HFS350EV) and maintained at a controllable temperature in the range of 80-295 K. The laser beam was focused on the sample surface using an infinity-corrected objective (Mitutoyo Plan APO 50x NIR with NA = 0.42) at normal incidence with a spot size of around 13 $\mu$m that provided the sample with uniform irradiation. The signal from a single nanocube was collected with the same objective and sent to a slit spectrometer coupled to a liquid-nitrogen-cooled imaging CCD camera (Princeton Instruments SP2500+PyLoN). To avoid parasitic signals from the neighborhood nanoparticles, a spatial filter was used.

\textbf{Numerical modeling}

We describe the dispersive permittivity of CsPbCl$_3$ by the two-pole Lorentz model, $\varepsilon (\omega) = \varepsilon_\infty + \sum_{n=1}^2 \left( \frac{A_n}{\omega-\Omega_n} - \frac{A_n^*}{\omega+\Omega_n^*} \right)$, where $\Omega_n$ and $A_n$ are complex constants~\cite{garcia2017extracting}. The unknown coefficients, presented in Supplementary materials, were found by approximating the experimental data on the permittivity of the CsPbCl$_3$ perovskite film at $80$~K. Figure~S3 of the Supplementary materials presents a comparison of the experimental and approximated dielectric function.
The substrate, consisting of $10$~nm Al$_2$O$_3$, $50$~nm Ag layer, and semi-infinite c-Si, was also taken into account in numerical calculations.
Ag layer was described by one Drude-Lorentz pole~\cite{yan2018rigorous} $\varepsilon (\omega) = \varepsilon_\infty - \frac{\varepsilon_\infty \omega_p^2}{\omega^2 - \omega_0^2 + i \omega \gamma}$, while c-Si by four Lorentz poles with the formula similar to that for the perovskite cube~\cite{garcia2017extracting}. The permittivity of Al$_2$O$_3$ was approximated by a constant value $\varepsilon = 2.89$.


To calculate the integral~\eqref{eq:PL_integral}, we applied the QNM theory~\cite{lalanne2018light, wu2023modal}. QNMs are source-free solutions of Maxwell's equations that were obtained using COMSOL eigensolver and normalized using the PML norm~\cite{sauvan2022normalization}. The QNM calculations were performed at two eigenvalue accumulation points: below and above the exciton frequency, around each of which 200 QNM were searched.

\textbf{Rate equations model}

We consider the joint dynamics of the discrete exciton-polariton population coupled to a momentum-resolved dark exciton reservoir.
For simplicity, we discard the backward process of reservoir repopulation from optically active states.
We also discard the scattering processes between reservoir states having different wave vectors.
Instead, we assume rapid thermalization of the reservoir, so that at each moment it is characterized by a Boltzmann-like distribution:
$n_k (t) =  n_{\rm R} (t) \times n_0 \exp \left[ -\left(\hbar^2 k^2 - \hbar^2 k_{\rm LC}^2) /  (2M_{\rm X} k_{\rm B} T \right) \right]$,
where $M_{\rm X}$ is the exciton total mass, $k_{\rm LC}$ is the exciton light cone edge defined as $E_{\rm X} +\hbar^2 k_{\rm LC}^2/(2M_{\rm X}) = \hbar c k_{\rm LC} / \sqrt{\varepsilon_{\rm s}} $, where $E_{\rm X}$ is the energy of exciton resonance, $\varepsilon_{\rm s}$ is the static dielectric constant.
The normalization constant is $n_0 \approx \left[ 2\pi \hbar^2 / (M_{\rm X} k_{\rm B} T) \right]^{3/2}$, provided that $k_{\rm LC}^3 \ll n_0$.
The systems of kinetic equations then have a form
\begin{align}
    \label{eq:dynamics_k_resolved}
    \frac{{\rm d} n_k }{{\rm d} t} &= - \frac{n_k}{\tau_{\rm R}}   
    -\sum_i \frac{n_k}{\tau_{k,i} }  (n_i +1)  
    + P_k e^{-(t-t_0)^2/(2\delta t^2)} , \\
    \frac{{\rm d} n_i }{{\rm d} t} &=  - \frac{n_i}{\tau_i}  
    +\sum_k \frac{n_k}{\tau_{k,i} }   (n_i +1) 
    + \sum_{j> i} \frac{1}{\tau_{j,i} } \left[  n_j (n_i +1) 
    - e^{-\hbar \omega_{\rm LO} / (k_{\rm B} T)} n_i (n_j +1) \right]  \notag \\
    & +\sum_{j< i} \frac{1}{\tau_{i,j} } \left[  e^{-\hbar \omega_{\rm LO} / (k_{\rm B} T)} n_j (n_i +1) 
    -  n_i (n_j +1) \right],
\end{align}
where $P_k = P_0 n_0 \exp \left[ -\left(\hbar^2 k^2 - \hbar^2 k_{\rm LC}^2) /  (2M_{\rm X} k_{\rm B} T \right) \right]$.
The exciton scattering rates on LO phonons can be quantified via Fermi golden rule as follows \cite{kavokin2017microcavities}:
\begin{align}
    \frac{1}{\tau_{k,i} } &= \frac{2\pi}{\hbar} |X_i|^2 |M(k)|^2 
    \left(\frac{1}{e^{\hbar \omega_{\rm LO} / (k_{\rm B} T)}-1} +1 \right) 
    \frac{\gamma_i/\pi}{ \left( E_{\rm X} + \hbar^2 k^2 /(2M_{\rm X}) - E_i -\hbar\omega_{\rm LO}  \right)^2 + (\gamma_i/\pi)^2}, \\
    \frac{1}{\tau_{j,i} } &= \frac{2\pi}{\hbar} |X_j|^2 |X_i|^2 |M(q_{\rm eff})|^2 
    \left(\frac{1}{e^{\hbar \omega_{\rm LO} / (k_{\rm B} T)}-1} +1 \right) 
    \frac{\gamma_i/\pi}{ \left( E_j - E_i -\hbar\omega_{\rm LO}  \right)^2 + (\gamma_i/\pi)^2},
\end{align}
where $X_i$ is the Hopfield coefficient characterizing exciton fraction in respective polariton state, $\gamma_i = \hbar / \tau_i$ is the decay rate,  
and the scattering matrix element 
\begin{align}
    M(q) = - \sqrt{\frac{e^2 \hbar\omega_{\rm LO} }{\varepsilon_0 \varepsilon^* L^3} } \frac{1}{q} .
\end{align}
Here $L$ is the lateral size of nanocuboid, 
and $1/\varepsilon^* = 1/ \varepsilon_\infty - 1/\varepsilon_{\rm s}$, 
with $\varepsilon_\infty$
denoting the high-frequency dielectric constant.
For the dispersive modes, the characteristic phonon wave vector is found from the momentum conservation rule within the scattering event.
Here, we estimate the wave vector scale via the lateral size of the nanocuboid, defining the wave vector quantization: $q_{\rm eff} =2\pi / L$.
Performing the summation over the wave vectors of Eqs. \eqref{eq:dynamics_k_resolved}, we end up in Eqs. \eqref{eq:dynamics} of the main text, where the effective scattering rates are
\begin{align}
    \frac{1}{\tau_{{\rm R},i}} = \frac{4\pi}{(2\pi)^3} n_0 \int\limits_{k_{\rm LC}}^\infty \frac{\exp \left[ -\left(\hbar^2 k^2 - \hbar^2 k_{\rm LC}^2) /  (2M_{\rm X} k_{\rm B} T \right) \right]}{\tau_{k,i}} k^2 {\rm d} k .
\end{align}
The material parameters for CsPbCl$_3$ are the following:
$E_{\rm X} = 3007$ meV, $\hbar\omega_{\rm LO} =26$ meV \cite{liao2019situ}, $\varepsilon_{\rm s} = 17.5$, $\varepsilon_\infty = 3.7$ \cite{filip2021phonon}.

\section*{Acknowledgements}
This work was supported by the National Natural Science Foundation of China (project 62350610272), the Priority 2030 Federal Academic Leadership Program, and the Department of Science and Technology of Shandong Province (Grant KY0020240040).
The authors are also thankful to Dr. Dolgintsev (ITMO) for assistance with SEM measurements and to Dr. Kirsanova (Skoltech) for assistance with STEM and EDX measurements.

\bibliography{pero_laser}

\begin{thebibliography}{44}%
\makeatletter
\providecommand \@ifxundefined [1]{%
 \@ifx{#1\undefined}
}%
\providecommand \@ifnum [1]{%
 \ifnum #1\expandafter \@firstoftwo
 \else \expandafter \@secondoftwo
 \fi
}%
\providecommand \@ifx [1]{%
 \ifx #1\expandafter \@firstoftwo
 \else \expandafter \@secondoftwo
 \fi
}%
\providecommand \natexlab [1]{#1}%
\providecommand \enquote  [1]{``#1''}%
\providecommand \bibnamefont  [1]{#1}%
\providecommand \bibfnamefont [1]{#1}%
\providecommand \citenamefont [1]{#1}%
\providecommand \href@noop [0]{\@secondoftwo}%
\providecommand \href [0]{\begingroup \@sanitize@url \@href}%
\providecommand \@href[1]{\@@startlink{#1}\@@href}%
\providecommand \@@href[1]{\endgroup#1\@@endlink}%
\providecommand \@sanitize@url [0]{\catcode `\\12\catcode `\$12\catcode `\&12\catcode `\#12\catcode `\^12\catcode `\_12\catcode `\%12\relax}%
\providecommand \@@startlink[1]{}%
\providecommand \@@endlink[0]{}%
\providecommand \url  [0]{\begingroup\@sanitize@url \@url }%
\providecommand \@url [1]{\endgroup\@href {#1}{\urlprefix }}%
\providecommand \urlprefix  [0]{URL }%
\providecommand \Eprint [0]{\href }%
\providecommand \doibase [0]{https://doi.org/}%
\providecommand \selectlanguage [0]{\@gobble}%
\providecommand \bibinfo  [0]{\@secondoftwo}%
\providecommand \bibfield  [0]{\@secondoftwo}%
\providecommand \translation [1]{[#1]}%
\providecommand \BibitemOpen [0]{}%
\providecommand \bibitemStop [0]{}%
\providecommand \bibitemNoStop [0]{.\EOS\space}%
\providecommand \EOS [0]{\spacefactor3000\relax}%
\providecommand \BibitemShut  [1]{\csname bibitem#1\endcsname}%
\let\auto@bib@innerbib\@empty
\bibitem [{\citenamefont {Ma}\ and\ \citenamefont {Oulton}(2019)}]{ma2019applications}%
  \BibitemOpen
  \bibfield  {author} {\bibinfo {author} {\bibfnamefont {R.-M.}\ \bibnamefont {Ma}}\ and\ \bibinfo {author} {\bibfnamefont {R.~F.}\ \bibnamefont {Oulton}},\ }\bibfield  {title} {\bibinfo {title} {Applications of nanolasers},\ }\href@noop {} {\bibfield  {journal} {\bibinfo  {journal} {Nature nanotechnology}\ }\textbf {\bibinfo {volume} {14}},\ \bibinfo {pages} {12} (\bibinfo {year} {2019})}\BibitemShut {NoStop}%
\bibitem [{\citenamefont {Azzam}\ \emph {et~al.}(2020)\citenamefont {Azzam}, \citenamefont {Kildishev}, \citenamefont {Ma}, \citenamefont {Ning}, \citenamefont {Oulton}, \citenamefont {Shalaev}, \citenamefont {Stockman}, \citenamefont {Xu},\ and\ \citenamefont {Zhang}}]{azzam2020ten}%
  \BibitemOpen
  \bibfield  {author} {\bibinfo {author} {\bibfnamefont {S.~I.}\ \bibnamefont {Azzam}}, \bibinfo {author} {\bibfnamefont {A.~V.}\ \bibnamefont {Kildishev}}, \bibinfo {author} {\bibfnamefont {R.-M.}\ \bibnamefont {Ma}}, \bibinfo {author} {\bibfnamefont {C.-Z.}\ \bibnamefont {Ning}}, \bibinfo {author} {\bibfnamefont {R.}~\bibnamefont {Oulton}}, \bibinfo {author} {\bibfnamefont {V.~M.}\ \bibnamefont {Shalaev}}, \bibinfo {author} {\bibfnamefont {M.~I.}\ \bibnamefont {Stockman}}, \bibinfo {author} {\bibfnamefont {J.-L.}\ \bibnamefont {Xu}},\ and\ \bibinfo {author} {\bibfnamefont {X.}~\bibnamefont {Zhang}},\ }\bibfield  {title} {\bibinfo {title} {Ten years of spasers and plasmonic nanolasers},\ }\href@noop {} {\bibfield  {journal} {\bibinfo  {journal} {Light: Science \& Applications}\ }\textbf {\bibinfo {volume} {9}},\ \bibinfo {pages} {90} (\bibinfo {year} {2020})}\BibitemShut {NoStop}%
\bibitem [{\citenamefont {Grade{\v{c}}ak}\ \emph {et~al.}(2005)\citenamefont {Grade{\v{c}}ak}, \citenamefont {Qian}, \citenamefont {Li}, \citenamefont {Park},\ and\ \citenamefont {Lieber}}]{gradevcak2005gan}%
  \BibitemOpen
  \bibfield  {author} {\bibinfo {author} {\bibfnamefont {S.}~\bibnamefont {Grade{\v{c}}ak}}, \bibinfo {author} {\bibfnamefont {F.}~\bibnamefont {Qian}}, \bibinfo {author} {\bibfnamefont {Y.}~\bibnamefont {Li}}, \bibinfo {author} {\bibfnamefont {H.-G.}\ \bibnamefont {Park}},\ and\ \bibinfo {author} {\bibfnamefont {C.~M.}\ \bibnamefont {Lieber}},\ }\bibfield  {title} {\bibinfo {title} {Gan nanowire lasers with low lasing thresholds},\ }\href@noop {} {\bibfield  {journal} {\bibinfo  {journal} {Applied Physics Letters}\ }\textbf {\bibinfo {volume} {87}},\ \bibinfo {pages} {173111} (\bibinfo {year} {2005})}\BibitemShut {NoStop}%
\bibitem [{\citenamefont {Zhang}\ \emph {et~al.}(2014)\citenamefont {Zhang}, \citenamefont {Li}, \citenamefont {Liu}, \citenamefont {Qian}, \citenamefont {Li}, \citenamefont {Sum}, \citenamefont {Lieber},\ and\ \citenamefont {Xiong}}]{zhang2014room}%
  \BibitemOpen
  \bibfield  {author} {\bibinfo {author} {\bibfnamefont {Q.}~\bibnamefont {Zhang}}, \bibinfo {author} {\bibfnamefont {G.}~\bibnamefont {Li}}, \bibinfo {author} {\bibfnamefont {X.}~\bibnamefont {Liu}}, \bibinfo {author} {\bibfnamefont {F.}~\bibnamefont {Qian}}, \bibinfo {author} {\bibfnamefont {Y.}~\bibnamefont {Li}}, \bibinfo {author} {\bibfnamefont {T.~C.}\ \bibnamefont {Sum}}, \bibinfo {author} {\bibfnamefont {C.~M.}\ \bibnamefont {Lieber}},\ and\ \bibinfo {author} {\bibfnamefont {Q.}~\bibnamefont {Xiong}},\ }\bibfield  {title} {\bibinfo {title} {A room temperature low-threshold ultraviolet plasmonic nanolaser},\ }\href@noop {} {\bibfield  {journal} {\bibinfo  {journal} {Nature communications}\ }\textbf {\bibinfo {volume} {5}},\ \bibinfo {pages} {4953} (\bibinfo {year} {2014})}\BibitemShut {NoStop}%
\bibitem [{\citenamefont {Huang}\ \emph {et~al.}(2001)\citenamefont {Huang}, \citenamefont {Mao}, \citenamefont {Feick}, \citenamefont {Yan}, \citenamefont {Wu}, \citenamefont {Kind}, \citenamefont {Weber}, \citenamefont {Russo},\ and\ \citenamefont {Yang}}]{huang2001room}%
  \BibitemOpen
  \bibfield  {author} {\bibinfo {author} {\bibfnamefont {M.~H.}\ \bibnamefont {Huang}}, \bibinfo {author} {\bibfnamefont {S.}~\bibnamefont {Mao}}, \bibinfo {author} {\bibfnamefont {H.}~\bibnamefont {Feick}}, \bibinfo {author} {\bibfnamefont {H.}~\bibnamefont {Yan}}, \bibinfo {author} {\bibfnamefont {Y.}~\bibnamefont {Wu}}, \bibinfo {author} {\bibfnamefont {H.}~\bibnamefont {Kind}}, \bibinfo {author} {\bibfnamefont {E.}~\bibnamefont {Weber}}, \bibinfo {author} {\bibfnamefont {R.}~\bibnamefont {Russo}},\ and\ \bibinfo {author} {\bibfnamefont {P.}~\bibnamefont {Yang}},\ }\bibfield  {title} {\bibinfo {title} {Room-temperature ultraviolet nanowire nanolasers},\ }\href@noop {} {\bibfield  {journal} {\bibinfo  {journal} {science}\ }\textbf {\bibinfo {volume} {292}},\ \bibinfo {pages} {1897} (\bibinfo {year} {2001})}\BibitemShut {NoStop}%
\bibitem [{\citenamefont {Duan}\ \emph {et~al.}(2003)\citenamefont {Duan}, \citenamefont {Huang}, \citenamefont {Agarwal},\ and\ \citenamefont {Lieber}}]{duan2003single}%
  \BibitemOpen
  \bibfield  {author} {\bibinfo {author} {\bibfnamefont {X.}~\bibnamefont {Duan}}, \bibinfo {author} {\bibfnamefont {Y.}~\bibnamefont {Huang}}, \bibinfo {author} {\bibfnamefont {R.}~\bibnamefont {Agarwal}},\ and\ \bibinfo {author} {\bibfnamefont {C.~M.}\ \bibnamefont {Lieber}},\ }\bibfield  {title} {\bibinfo {title} {Single-nanowire electrically driven lasers},\ }\href@noop {} {\bibfield  {journal} {\bibinfo  {journal} {Nature}\ }\textbf {\bibinfo {volume} {421}},\ \bibinfo {pages} {241} (\bibinfo {year} {2003})}\BibitemShut {NoStop}%
\bibitem [{\citenamefont {Oulton}\ \emph {et~al.}(2009)\citenamefont {Oulton}, \citenamefont {Sorger}, \citenamefont {Zentgraf}, \citenamefont {Ma}, \citenamefont {Gladden}, \citenamefont {Dai}, \citenamefont {Bartal},\ and\ \citenamefont {Zhang}}]{oulton2009plasmon}%
  \BibitemOpen
  \bibfield  {author} {\bibinfo {author} {\bibfnamefont {R.~F.}\ \bibnamefont {Oulton}}, \bibinfo {author} {\bibfnamefont {V.~J.}\ \bibnamefont {Sorger}}, \bibinfo {author} {\bibfnamefont {T.}~\bibnamefont {Zentgraf}}, \bibinfo {author} {\bibfnamefont {R.-M.}\ \bibnamefont {Ma}}, \bibinfo {author} {\bibfnamefont {C.}~\bibnamefont {Gladden}}, \bibinfo {author} {\bibfnamefont {L.}~\bibnamefont {Dai}}, \bibinfo {author} {\bibfnamefont {G.}~\bibnamefont {Bartal}},\ and\ \bibinfo {author} {\bibfnamefont {X.}~\bibnamefont {Zhang}},\ }\bibfield  {title} {\bibinfo {title} {Plasmon lasers at deep subwavelength scale},\ }\href@noop {} {\bibfield  {journal} {\bibinfo  {journal} {nature}\ }\textbf {\bibinfo {volume} {461}},\ \bibinfo {pages} {629} (\bibinfo {year} {2009})}\BibitemShut {NoStop}%
\bibitem [{\citenamefont {Zhuang}\ \emph {et~al.}(2012)\citenamefont {Zhuang}, \citenamefont {Ning},\ and\ \citenamefont {Pan}}]{zhuang2012composition}%
  \BibitemOpen
  \bibfield  {author} {\bibinfo {author} {\bibfnamefont {X.}~\bibnamefont {Zhuang}}, \bibinfo {author} {\bibfnamefont {C.-Z.}\ \bibnamefont {Ning}},\ and\ \bibinfo {author} {\bibfnamefont {A.}~\bibnamefont {Pan}},\ }\bibfield  {title} {\bibinfo {title} {Composition and bandgap-graded semiconductor alloy nanowires},\ }\href@noop {} {\bibfield  {journal} {\bibinfo  {journal} {Advanced Materials}\ }\textbf {\bibinfo {volume} {24}},\ \bibinfo {pages} {13} (\bibinfo {year} {2012})}\BibitemShut {NoStop}%
\bibitem [{\citenamefont {Chang}\ \emph {et~al.}(2021)\citenamefont {Chang}, \citenamefont {Zuo}, \citenamefont {Liu}, \citenamefont {Ou}, \citenamefont {Ding}, \citenamefont {Yang}, \citenamefont {Feng}, \citenamefont {Cao}, \citenamefont {Lin}, \citenamefont {Ren} \emph {et~al.}}]{chang2021tuning}%
  \BibitemOpen
  \bibfield  {author} {\bibinfo {author} {\bibfnamefont {Y.-Z.}\ \bibnamefont {Chang}}, \bibinfo {author} {\bibfnamefont {Z.-Y.}\ \bibnamefont {Zuo}}, \bibinfo {author} {\bibfnamefont {Y.-Y.}\ \bibnamefont {Liu}}, \bibinfo {author} {\bibfnamefont {C.-J.}\ \bibnamefont {Ou}}, \bibinfo {author} {\bibfnamefont {X.-H.}\ \bibnamefont {Ding}}, \bibinfo {author} {\bibfnamefont {L.}~\bibnamefont {Yang}}, \bibinfo {author} {\bibfnamefont {Q.-Y.}\ \bibnamefont {Feng}}, \bibinfo {author} {\bibfnamefont {H.-T.}\ \bibnamefont {Cao}}, \bibinfo {author} {\bibfnamefont {J.-Y.}\ \bibnamefont {Lin}}, \bibinfo {author} {\bibfnamefont {B.-Y.}\ \bibnamefont {Ren}}, \emph {et~al.},\ }\bibfield  {title} {\bibinfo {title} {Tuning stimulated emission properties of oligofluorene-based gain media via non-conjugation strategy},\ }\href@noop {} {\bibfield  {journal} {\bibinfo  {journal} {Dyes and Pigments}\ }\textbf {\bibinfo {volume} {186}},\ \bibinfo {pages} {109037} (\bibinfo {year} {2021})}\BibitemShut {NoStop}%
\bibitem [{\citenamefont {Tatarinov}\ \emph {et~al.}(2023)\citenamefont {Tatarinov}, \citenamefont {Anoshkin}, \citenamefont {Tsibizov}, \citenamefont {Sheremet}, \citenamefont {Isik}, \citenamefont {Zhizhchenko}, \citenamefont {Cherepakhin}, \citenamefont {Kuchmizhak}, \citenamefont {Pushkarev}, \citenamefont {Demir} \emph {et~al.}}]{tatarinov2023high}%
  \BibitemOpen
  \bibfield  {author} {\bibinfo {author} {\bibfnamefont {D.~A.}\ \bibnamefont {Tatarinov}}, \bibinfo {author} {\bibfnamefont {S.~S.}\ \bibnamefont {Anoshkin}}, \bibinfo {author} {\bibfnamefont {I.~A.}\ \bibnamefont {Tsibizov}}, \bibinfo {author} {\bibfnamefont {V.}~\bibnamefont {Sheremet}}, \bibinfo {author} {\bibfnamefont {F.}~\bibnamefont {Isik}}, \bibinfo {author} {\bibfnamefont {A.~Y.}\ \bibnamefont {Zhizhchenko}}, \bibinfo {author} {\bibfnamefont {A.~B.}\ \bibnamefont {Cherepakhin}}, \bibinfo {author} {\bibfnamefont {A.~A.}\ \bibnamefont {Kuchmizhak}}, \bibinfo {author} {\bibfnamefont {A.~P.}\ \bibnamefont {Pushkarev}}, \bibinfo {author} {\bibfnamefont {H.~V.}\ \bibnamefont {Demir}}, \emph {et~al.},\ }\bibfield  {title} {\bibinfo {title} {High-quality cspbbr3 perovskite films with modal gain above 10 000 cm- 1 at room temperature},\ }\href@noop {} {\bibfield  {journal} {\bibinfo  {journal} {Advanced Optical Materials}\ }\textbf {\bibinfo {volume} {11}},\ \bibinfo {pages} {2202407} (\bibinfo {year}
  {2023})}\BibitemShut {NoStop}%
\bibitem [{\citenamefont {Shi}\ \emph {et~al.}(2025)\citenamefont {Shi}, \citenamefont {Deng}, \citenamefont {Gan}, \citenamefont {Xu}, \citenamefont {Zhang},\ and\ \citenamefont {Xiong}}]{shi2025ten}%
  \BibitemOpen
  \bibfield  {author} {\bibinfo {author} {\bibfnamefont {Y.}~\bibnamefont {Shi}}, \bibinfo {author} {\bibfnamefont {X.}~\bibnamefont {Deng}}, \bibinfo {author} {\bibfnamefont {Y.}~\bibnamefont {Gan}}, \bibinfo {author} {\bibfnamefont {L.}~\bibnamefont {Xu}}, \bibinfo {author} {\bibfnamefont {Q.}~\bibnamefont {Zhang}},\ and\ \bibinfo {author} {\bibfnamefont {Q.}~\bibnamefont {Xiong}},\ }\bibfield  {title} {\bibinfo {title} {Ten years of perovskite lasers},\ }\href@noop {} {\bibfield  {journal} {\bibinfo  {journal} {Advanced Materials}\ ,\ \bibinfo {pages} {2413559}} (\bibinfo {year} {2025})}\BibitemShut {NoStop}%
\bibitem [{\citenamefont {Wang}\ \emph {et~al.}(2020)\citenamefont {Wang}, \citenamefont {Han}, \citenamefont {Wang}, \citenamefont {Yang}, \citenamefont {Liu}, \citenamefont {Huang}, \citenamefont {Zhang}, \citenamefont {Chang}, \citenamefont {Wu},\ and\ \citenamefont {Zhong}}]{wang2020dimension}%
  \BibitemOpen
  \bibfield  {author} {\bibinfo {author} {\bibfnamefont {C.}~\bibnamefont {Wang}}, \bibinfo {author} {\bibfnamefont {D.}~\bibnamefont {Han}}, \bibinfo {author} {\bibfnamefont {J.}~\bibnamefont {Wang}}, \bibinfo {author} {\bibfnamefont {Y.}~\bibnamefont {Yang}}, \bibinfo {author} {\bibfnamefont {X.}~\bibnamefont {Liu}}, \bibinfo {author} {\bibfnamefont {S.}~\bibnamefont {Huang}}, \bibinfo {author} {\bibfnamefont {X.}~\bibnamefont {Zhang}}, \bibinfo {author} {\bibfnamefont {S.}~\bibnamefont {Chang}}, \bibinfo {author} {\bibfnamefont {K.}~\bibnamefont {Wu}},\ and\ \bibinfo {author} {\bibfnamefont {H.}~\bibnamefont {Zhong}},\ }\bibfield  {title} {\bibinfo {title} {Dimension control of in situ fabricated cspbclbr2 nanocrystal films toward efficient blue light-emitting diodes},\ }\href@noop {} {\bibfield  {journal} {\bibinfo  {journal} {Nature Communications}\ }\textbf {\bibinfo {volume} {11}},\ \bibinfo {pages} {6428} (\bibinfo {year} {2020})}\BibitemShut {NoStop}%
\bibitem [{\citenamefont {Karlsson}\ \emph {et~al.}(2021)\citenamefont {Karlsson}, \citenamefont {Yi}, \citenamefont {Reichert}, \citenamefont {Luo}, \citenamefont {Lin}, \citenamefont {Zhang}, \citenamefont {Bao}, \citenamefont {Zhang}, \citenamefont {Bai}, \citenamefont {Zheng} \emph {et~al.}}]{karlsson2021mixed}%
  \BibitemOpen
  \bibfield  {author} {\bibinfo {author} {\bibfnamefont {M.}~\bibnamefont {Karlsson}}, \bibinfo {author} {\bibfnamefont {Z.}~\bibnamefont {Yi}}, \bibinfo {author} {\bibfnamefont {S.}~\bibnamefont {Reichert}}, \bibinfo {author} {\bibfnamefont {X.}~\bibnamefont {Luo}}, \bibinfo {author} {\bibfnamefont {W.}~\bibnamefont {Lin}}, \bibinfo {author} {\bibfnamefont {Z.}~\bibnamefont {Zhang}}, \bibinfo {author} {\bibfnamefont {C.}~\bibnamefont {Bao}}, \bibinfo {author} {\bibfnamefont {R.}~\bibnamefont {Zhang}}, \bibinfo {author} {\bibfnamefont {S.}~\bibnamefont {Bai}}, \bibinfo {author} {\bibfnamefont {G.}~\bibnamefont {Zheng}}, \emph {et~al.},\ }\bibfield  {title} {\bibinfo {title} {Mixed halide perovskites for spectrally stable and high-efficiency blue light-emitting diodes},\ }\href@noop {} {\bibfield  {journal} {\bibinfo  {journal} {Nature Communications}\ }\textbf {\bibinfo {volume} {12}},\ \bibinfo {pages} {361} (\bibinfo {year} {2021})}\BibitemShut {NoStop}%
\bibitem [{\citenamefont {Liu}\ \emph {et~al.}(2021)\citenamefont {Liu}, \citenamefont {Zhang}, \citenamefont {Chen}, \citenamefont {Liu}, \citenamefont {Li}, \citenamefont {Wu}, \citenamefont {Wang}, \citenamefont {Jiang}, \citenamefont {Li}, \citenamefont {Li} \emph {et~al.}}]{liu2021water}%
  \BibitemOpen
  \bibfield  {author} {\bibinfo {author} {\bibfnamefont {Y.}~\bibnamefont {Liu}}, \bibinfo {author} {\bibfnamefont {L.}~\bibnamefont {Zhang}}, \bibinfo {author} {\bibfnamefont {S.}~\bibnamefont {Chen}}, \bibinfo {author} {\bibfnamefont {C.}~\bibnamefont {Liu}}, \bibinfo {author} {\bibfnamefont {Y.}~\bibnamefont {Li}}, \bibinfo {author} {\bibfnamefont {J.}~\bibnamefont {Wu}}, \bibinfo {author} {\bibfnamefont {D.}~\bibnamefont {Wang}}, \bibinfo {author} {\bibfnamefont {Z.}~\bibnamefont {Jiang}}, \bibinfo {author} {\bibfnamefont {Y.}~\bibnamefont {Li}}, \bibinfo {author} {\bibfnamefont {Y.}~\bibnamefont {Li}}, \emph {et~al.},\ }\bibfield  {title} {\bibinfo {title} {Water-soluble conjugated polyelectrolyte hole transporting layer for efficient sky-blue perovskite light-emitting diodes},\ }\href@noop {} {\bibfield  {journal} {\bibinfo  {journal} {Small}\ }\textbf {\bibinfo {volume} {17}},\ \bibinfo {pages} {2101477} (\bibinfo {year} {2021})}\BibitemShut {NoStop}%
\bibitem [{\citenamefont {Sun}\ \emph {et~al.}(2022)\citenamefont {Sun}, \citenamefont {Lu}, \citenamefont {Zhong}, \citenamefont {Lu}, \citenamefont {Qin}, \citenamefont {Gao}, \citenamefont {Bai}, \citenamefont {Wu},\ and\ \citenamefont {Zhang}}]{sun2022bifunctional}%
  \BibitemOpen
  \bibfield  {author} {\bibinfo {author} {\bibfnamefont {S.}~\bibnamefont {Sun}}, \bibinfo {author} {\bibfnamefont {M.}~\bibnamefont {Lu}}, \bibinfo {author} {\bibfnamefont {Y.}~\bibnamefont {Zhong}}, \bibinfo {author} {\bibfnamefont {P.}~\bibnamefont {Lu}}, \bibinfo {author} {\bibfnamefont {F.}~\bibnamefont {Qin}}, \bibinfo {author} {\bibfnamefont {Y.}~\bibnamefont {Gao}}, \bibinfo {author} {\bibfnamefont {X.}~\bibnamefont {Bai}}, \bibinfo {author} {\bibfnamefont {Z.}~\bibnamefont {Wu}},\ and\ \bibinfo {author} {\bibfnamefont {Y.}~\bibnamefont {Zhang}},\ }\bibfield  {title} {\bibinfo {title} {Bifunctional molecule enables high-quality cspb (br/cl) 3 nanocrystals for efficient and stable pure-blue perovskite light-emitting diodes},\ }\href@noop {} {\bibfield  {journal} {\bibinfo  {journal} {ACS Energy Letters}\ }\textbf {\bibinfo {volume} {7}},\ \bibinfo {pages} {3974} (\bibinfo {year} {2022})}\BibitemShut {NoStop}%
\bibitem [{\citenamefont {Baranowski}\ \emph {et~al.}(2020)\citenamefont {Baranowski}, \citenamefont {Plochocka}, \citenamefont {Su}, \citenamefont {Legrand}, \citenamefont {Barisien}, \citenamefont {Bernardot}, \citenamefont {Xiong}, \citenamefont {Testelin},\ and\ \citenamefont {Chamarro}}]{baranowski2020exciton}%
  \BibitemOpen
  \bibfield  {author} {\bibinfo {author} {\bibfnamefont {M.}~\bibnamefont {Baranowski}}, \bibinfo {author} {\bibfnamefont {P.}~\bibnamefont {Plochocka}}, \bibinfo {author} {\bibfnamefont {R.}~\bibnamefont {Su}}, \bibinfo {author} {\bibfnamefont {L.}~\bibnamefont {Legrand}}, \bibinfo {author} {\bibfnamefont {T.}~\bibnamefont {Barisien}}, \bibinfo {author} {\bibfnamefont {F.}~\bibnamefont {Bernardot}}, \bibinfo {author} {\bibfnamefont {Q.}~\bibnamefont {Xiong}}, \bibinfo {author} {\bibfnamefont {C.}~\bibnamefont {Testelin}},\ and\ \bibinfo {author} {\bibfnamefont {M.}~\bibnamefont {Chamarro}},\ }\bibfield  {title} {\bibinfo {title} {Exciton binding energy and effective mass of cspbcl\_3: a magneto-optical study},\ }\href@noop {} {\bibfield  {journal} {\bibinfo  {journal} {Photonics Research}\ }\textbf {\bibinfo {volume} {8}},\ \bibinfo {pages} {A50} (\bibinfo {year} {2020})}\BibitemShut {NoStop}%
\bibitem [{\citenamefont {Masharin}\ \emph {et~al.}(2024)\citenamefont {Masharin}, \citenamefont {Khmelevskaia}, \citenamefont {Kondratiev}, \citenamefont {Markina}, \citenamefont {Utyushev}, \citenamefont {Dolgintsev}, \citenamefont {Dmitriev}, \citenamefont {Shahnazaryan}, \citenamefont {Pushkarev}, \citenamefont {Isik} \emph {et~al.}}]{masharin2024polariton}%
  \BibitemOpen
  \bibfield  {author} {\bibinfo {author} {\bibfnamefont {M.~A.}\ \bibnamefont {Masharin}}, \bibinfo {author} {\bibfnamefont {D.}~\bibnamefont {Khmelevskaia}}, \bibinfo {author} {\bibfnamefont {V.~I.}\ \bibnamefont {Kondratiev}}, \bibinfo {author} {\bibfnamefont {D.~I.}\ \bibnamefont {Markina}}, \bibinfo {author} {\bibfnamefont {A.~D.}\ \bibnamefont {Utyushev}}, \bibinfo {author} {\bibfnamefont {D.~M.}\ \bibnamefont {Dolgintsev}}, \bibinfo {author} {\bibfnamefont {A.~D.}\ \bibnamefont {Dmitriev}}, \bibinfo {author} {\bibfnamefont {V.~A.}\ \bibnamefont {Shahnazaryan}}, \bibinfo {author} {\bibfnamefont {A.~P.}\ \bibnamefont {Pushkarev}}, \bibinfo {author} {\bibfnamefont {F.}~\bibnamefont {Isik}}, \emph {et~al.},\ }\bibfield  {title} {\bibinfo {title} {Polariton lasing in mie-resonant perovskite nanocavity},\ }\href@noop {} {\bibfield  {journal} {\bibinfo  {journal} {Opto-Electron Adv 7, 230148 (2024)}\ }\textbf {\bibinfo {volume} {7}},\ \bibinfo {pages} {230148} (\bibinfo {year} {2024})}\BibitemShut {NoStop}%
\bibitem [{\citenamefont {Ermolaev}\ \emph {et~al.}(2023)\citenamefont {Ermolaev}, \citenamefont {Pushkarev}, \citenamefont {Zhizhchenko}, \citenamefont {Kuchmizhak}, \citenamefont {Iorsh}, \citenamefont {Kruglov}, \citenamefont {Mazitov}, \citenamefont {Ishteev}, \citenamefont {Konstantinova}, \citenamefont {Saranin} \emph {et~al.}}]{ermolaev2023giant}%
  \BibitemOpen
  \bibfield  {author} {\bibinfo {author} {\bibfnamefont {G.}~\bibnamefont {Ermolaev}}, \bibinfo {author} {\bibfnamefont {A.~P.}\ \bibnamefont {Pushkarev}}, \bibinfo {author} {\bibfnamefont {A.}~\bibnamefont {Zhizhchenko}}, \bibinfo {author} {\bibfnamefont {A.~A.}\ \bibnamefont {Kuchmizhak}}, \bibinfo {author} {\bibfnamefont {I.}~\bibnamefont {Iorsh}}, \bibinfo {author} {\bibfnamefont {I.}~\bibnamefont {Kruglov}}, \bibinfo {author} {\bibfnamefont {A.}~\bibnamefont {Mazitov}}, \bibinfo {author} {\bibfnamefont {A.}~\bibnamefont {Ishteev}}, \bibinfo {author} {\bibfnamefont {K.}~\bibnamefont {Konstantinova}}, \bibinfo {author} {\bibfnamefont {D.}~\bibnamefont {Saranin}}, \emph {et~al.},\ }\bibfield  {title} {\bibinfo {title} {Giant and tunable excitonic optical anisotropy in single-crystal halide perovskites},\ }\href@noop {} {\bibfield  {journal} {\bibinfo  {journal} {Nano Letters}\ }\textbf {\bibinfo {volume} {23}},\ \bibinfo {pages} {2570} (\bibinfo {year} {2023})}\BibitemShut {NoStop}%
\bibitem [{\citenamefont {Sauvan}\ \emph {et~al.}(2013)\citenamefont {Sauvan}, \citenamefont {Hugonin}, \citenamefont {Maksymov},\ and\ \citenamefont {Lalanne}}]{Sauvan2013}%
  \BibitemOpen
  \bibfield  {author} {\bibinfo {author} {\bibfnamefont {C.}~\bibnamefont {Sauvan}}, \bibinfo {author} {\bibfnamefont {J.~P.}\ \bibnamefont {Hugonin}}, \bibinfo {author} {\bibfnamefont {I.~S.}\ \bibnamefont {Maksymov}},\ and\ \bibinfo {author} {\bibfnamefont {P.}~\bibnamefont {Lalanne}},\ }\bibfield  {title} {\bibinfo {title} {{Theory of the spontaneous optical emission of nanosize photonic and plasmon resonators}},\ }\href@noop {} {\bibfield  {journal} {\bibinfo  {journal} {Physical Review Letters}\ }\textbf {\bibinfo {volume} {110}},\ \bibinfo {pages} {237401} (\bibinfo {year} {2013})}\BibitemShut {NoStop}%
\bibitem [{\citenamefont {Lalanne}\ \emph {et~al.}(2018)\citenamefont {Lalanne}, \citenamefont {Yan}, \citenamefont {Vynck}, \citenamefont {Sauvan},\ and\ \citenamefont {Hugonin}}]{lalanne2018light}%
  \BibitemOpen
  \bibfield  {author} {\bibinfo {author} {\bibfnamefont {P.}~\bibnamefont {Lalanne}}, \bibinfo {author} {\bibfnamefont {W.}~\bibnamefont {Yan}}, \bibinfo {author} {\bibfnamefont {K.}~\bibnamefont {Vynck}}, \bibinfo {author} {\bibfnamefont {C.}~\bibnamefont {Sauvan}},\ and\ \bibinfo {author} {\bibfnamefont {J.-P.}\ \bibnamefont {Hugonin}},\ }\bibfield  {title} {\bibinfo {title} {Light interaction with photonic and plasmonic resonances},\ }\href@noop {} {\bibfield  {journal} {\bibinfo  {journal} {Laser \& Photonics Reviews}\ }\textbf {\bibinfo {volume} {12}},\ \bibinfo {pages} {1700113} (\bibinfo {year} {2018})}\BibitemShut {NoStop}%
\bibitem [{\citenamefont {Yan}\ \emph {et~al.}(2018)\citenamefont {Yan}, \citenamefont {Faggiani},\ and\ \citenamefont {Lalanne}}]{yan2018rigorous}%
  \BibitemOpen
  \bibfield  {author} {\bibinfo {author} {\bibfnamefont {W.}~\bibnamefont {Yan}}, \bibinfo {author} {\bibfnamefont {R.}~\bibnamefont {Faggiani}},\ and\ \bibinfo {author} {\bibfnamefont {P.}~\bibnamefont {Lalanne}},\ }\bibfield  {title} {\bibinfo {title} {Rigorous modal analysis of plasmonic nanoresonators},\ }\href@noop {} {\bibfield  {journal} {\bibinfo  {journal} {Physical Review B}\ }\textbf {\bibinfo {volume} {97}},\ \bibinfo {pages} {205422} (\bibinfo {year} {2018})}\BibitemShut {NoStop}%
\bibitem [{\citenamefont {Wu}\ \emph {et~al.}(2020)\citenamefont {Wu}, \citenamefont {Baron}, \citenamefont {Lalanne},\ and\ \citenamefont {Vynck}}]{wu2020intrinsic}%
  \BibitemOpen
  \bibfield  {author} {\bibinfo {author} {\bibfnamefont {T.}~\bibnamefont {Wu}}, \bibinfo {author} {\bibfnamefont {A.}~\bibnamefont {Baron}}, \bibinfo {author} {\bibfnamefont {P.}~\bibnamefont {Lalanne}},\ and\ \bibinfo {author} {\bibfnamefont {K.}~\bibnamefont {Vynck}},\ }\bibfield  {title} {\bibinfo {title} {Intrinsic multipolar contents of nanoresonators for tailored scattering},\ }\href@noop {} {\bibfield  {journal} {\bibinfo  {journal} {Physical Review A}\ }\textbf {\bibinfo {volume} {101}},\ \bibinfo {pages} {011803} (\bibinfo {year} {2020})}\BibitemShut {NoStop}%
\bibitem [{\citenamefont {Yoshle}\ \emph {et~al.}(2004)\citenamefont {Yoshle}, \citenamefont {Scherer}, \citenamefont {Hendrickson}, \citenamefont {Khitrova}, \citenamefont {Gibbs}, \citenamefont {Rupper}, \citenamefont {Ell}, \citenamefont {Shchekin}, \citenamefont {Deppe}, \citenamefont {Yoshie}, \citenamefont {Scherer}, \citenamefont {Hendrickson}, \citenamefont {Khitrova}, \citenamefont {Gibbs}, \citenamefont {Rupper}, \citenamefont {Ell}, \citenamefont {Shchekin},\ and\ \citenamefont {Deppe}}]{Yoshle2004}%
  \BibitemOpen
  \bibfield  {author} {\bibinfo {author} {\bibfnamefont {T.}~\bibnamefont {Yoshle}}, \bibinfo {author} {\bibfnamefont {A.}~\bibnamefont {Scherer}}, \bibinfo {author} {\bibfnamefont {J.}~\bibnamefont {Hendrickson}}, \bibinfo {author} {\bibfnamefont {G.}~\bibnamefont {Khitrova}}, \bibinfo {author} {\bibfnamefont {H.~M.}\ \bibnamefont {Gibbs}}, \bibinfo {author} {\bibfnamefont {G.}~\bibnamefont {Rupper}}, \bibinfo {author} {\bibfnamefont {C.}~\bibnamefont {Ell}}, \bibinfo {author} {\bibfnamefont {O.~B.}\ \bibnamefont {Shchekin}}, \bibinfo {author} {\bibfnamefont {D.~G.}\ \bibnamefont {Deppe}}, \bibinfo {author} {\bibfnamefont {T.}~\bibnamefont {Yoshie}}, \bibinfo {author} {\bibfnamefont {A.}~\bibnamefont {Scherer}}, \bibinfo {author} {\bibfnamefont {J.}~\bibnamefont {Hendrickson}}, \bibinfo {author} {\bibfnamefont {G.}~\bibnamefont {Khitrova}}, \bibinfo {author} {\bibfnamefont {H.~M.}\ \bibnamefont {Gibbs}}, \bibinfo {author} {\bibfnamefont {G.}~\bibnamefont {Rupper}}, \bibinfo {author} {\bibfnamefont
  {C.}~\bibnamefont {Ell}}, \bibinfo {author} {\bibfnamefont {O.~B.}\ \bibnamefont {Shchekin}},\ and\ \bibinfo {author} {\bibfnamefont {D.~G.}\ \bibnamefont {Deppe}},\ }\bibfield  {title} {\bibinfo {title} {{Vacuum Rabi splitting with a single quantum dot in a photonic crystal nanocavity}},\ }\href@noop {} {\bibfield  {journal} {\bibinfo  {journal} {Nature}\ }\textbf {\bibinfo {volume} {432}},\ \bibinfo {pages} {200} (\bibinfo {year} {2004})}\BibitemShut {NoStop}%
\bibitem [{\citenamefont {Reithmaier}\ \emph {et~al.}(2004)\citenamefont {Reithmaier}, \citenamefont {Sek}, \citenamefont {Loffler}, \citenamefont {Hofmann}, \citenamefont {Kuhn}, \citenamefont {Reitzenstein}, \citenamefont {Keldysh}, \citenamefont {Kulakovskii}, \citenamefont {Reinecke},\ and\ \citenamefont {Forchel}}]{Reithmaier2004}%
  \BibitemOpen
  \bibfield  {author} {\bibinfo {author} {\bibfnamefont {J.~P.}\ \bibnamefont {Reithmaier}}, \bibinfo {author} {\bibfnamefont {G.}~\bibnamefont {Sek}}, \bibinfo {author} {\bibfnamefont {A.}~\bibnamefont {Loffler}}, \bibinfo {author} {\bibfnamefont {C.}~\bibnamefont {Hofmann}}, \bibinfo {author} {\bibfnamefont {S.}~\bibnamefont {Kuhn}}, \bibinfo {author} {\bibfnamefont {S.}~\bibnamefont {Reitzenstein}}, \bibinfo {author} {\bibfnamefont {L.~V.}\ \bibnamefont {Keldysh}}, \bibinfo {author} {\bibfnamefont {V.~D.}\ \bibnamefont {Kulakovskii}}, \bibinfo {author} {\bibfnamefont {T.~L.}\ \bibnamefont {Reinecke}},\ and\ \bibinfo {author} {\bibfnamefont {A.}~\bibnamefont {Forchel}},\ }\bibfield  {title} {\bibinfo {title} {{Strong coupling in a single quantum dot – semiconductor microcavity system}},\ }\href@noop {} {\bibfield  {journal} {\bibinfo  {journal} {Nature}\ }\textbf {\bibinfo {volume} {432}},\ \bibinfo {pages} {197} (\bibinfo {year} {2004})}\BibitemShut {NoStop}%
\bibitem [{\citenamefont {Gro{\ss}}\ \emph {et~al.}(2018)\citenamefont {Gro{\ss}}, \citenamefont {Hamm}, \citenamefont {Tufarelli}, \citenamefont {Hess},\ and\ \citenamefont {Hecht}}]{Gross2018}%
  \BibitemOpen
  \bibfield  {author} {\bibinfo {author} {\bibfnamefont {H.}~\bibnamefont {Gro{\ss}}}, \bibinfo {author} {\bibfnamefont {J.~M.}\ \bibnamefont {Hamm}}, \bibinfo {author} {\bibfnamefont {T.}~\bibnamefont {Tufarelli}}, \bibinfo {author} {\bibfnamefont {O.}~\bibnamefont {Hess}},\ and\ \bibinfo {author} {\bibfnamefont {B.}~\bibnamefont {Hecht}},\ }\bibfield  {title} {\bibinfo {title} {{Near-field strong coupling of single quantum dots}},\ }\href@noop {} {\bibfield  {journal} {\bibinfo  {journal} {Science Advances}\ }\textbf {\bibinfo {volume} {4}},\ \bibinfo {pages} {eaar4906} (\bibinfo {year} {2018})}\BibitemShut {NoStop}%
\bibitem [{\citenamefont {Bellessa}\ \emph {et~al.}(2004)\citenamefont {Bellessa}, \citenamefont {Bonnand}, \citenamefont {Plenet},\ and\ \citenamefont {Mugnier}}]{Bellessa2004}%
  \BibitemOpen
  \bibfield  {author} {\bibinfo {author} {\bibfnamefont {J.}~\bibnamefont {Bellessa}}, \bibinfo {author} {\bibfnamefont {C.}~\bibnamefont {Bonnand}}, \bibinfo {author} {\bibfnamefont {J.~C.}\ \bibnamefont {Plenet}},\ and\ \bibinfo {author} {\bibfnamefont {J.}~\bibnamefont {Mugnier}},\ }\bibfield  {title} {\bibinfo {title} {{Strong coupling between surface plasmons and excitons in an organic semiconductor}},\ }\href@noop {} {\bibfield  {journal} {\bibinfo  {journal} {Physical Review Letters}\ }\textbf {\bibinfo {volume} {93}},\ \bibinfo {pages} {36404} (\bibinfo {year} {2004})}\BibitemShut {NoStop}%
\bibitem [{\citenamefont {Wersall}\ \emph {et~al.}(2017)\citenamefont {Wersall}, \citenamefont {Cuadra}, \citenamefont {Antosiewicz}, \citenamefont {Balci},\ and\ \citenamefont {Shegai}}]{Wersall2017}%
  \BibitemOpen
  \bibfield  {author} {\bibinfo {author} {\bibfnamefont {M.}~\bibnamefont {Wersall}}, \bibinfo {author} {\bibfnamefont {J.}~\bibnamefont {Cuadra}}, \bibinfo {author} {\bibfnamefont {T.~J.}\ \bibnamefont {Antosiewicz}}, \bibinfo {author} {\bibfnamefont {S.}~\bibnamefont {Balci}},\ and\ \bibinfo {author} {\bibfnamefont {T.}~\bibnamefont {Shegai}},\ }\bibfield  {title} {\bibinfo {title} {{Observation of mode splitting in photoluminescence of individual plasmonic nanoparticles strongly coupled to molecular excitons}},\ }\href@noop {} {\bibfield  {journal} {\bibinfo  {journal} {Nano Letters}\ }\textbf {\bibinfo {volume} {17}},\ \bibinfo {pages} {551} (\bibinfo {year} {2017})}\BibitemShut {NoStop}%
\bibitem [{\citenamefont {Wers{\"{a}}ll}\ \emph {et~al.}(2019)\citenamefont {Wers{\"{a}}ll}, \citenamefont {Munkhbat}, \citenamefont {Baranov}, \citenamefont {Herrera}, \citenamefont {Cao}, \citenamefont {Antosiewicz},\ and\ \citenamefont {Shegai}}]{Wersaell2019}%
  \BibitemOpen
  \bibfield  {author} {\bibinfo {author} {\bibfnamefont {M.}~\bibnamefont {Wers{\"{a}}ll}}, \bibinfo {author} {\bibfnamefont {B.}~\bibnamefont {Munkhbat}}, \bibinfo {author} {\bibfnamefont {D.~G.}\ \bibnamefont {Baranov}}, \bibinfo {author} {\bibfnamefont {F.}~\bibnamefont {Herrera}}, \bibinfo {author} {\bibfnamefont {J.}~\bibnamefont {Cao}}, \bibinfo {author} {\bibfnamefont {T.~J.}\ \bibnamefont {Antosiewicz}},\ and\ \bibinfo {author} {\bibfnamefont {T.}~\bibnamefont {Shegai}},\ }\bibfield  {title} {\bibinfo {title} {{Correlative Dark-Field and Photoluminescence Spectroscopy of Individual Plasmon-Molecule Hybrid Nanostructures in a Strong Coupling Regime}},\ }\href@noop {} {\bibfield  {journal} {\bibinfo  {journal} {ACS Photonics}\ }\textbf {\bibinfo {volume} {6}},\ \bibinfo {pages} {2570} (\bibinfo {year} {2019})}\BibitemShut {NoStop}%
\bibitem [{\citenamefont {Shang}\ \emph {et~al.}(2018)\citenamefont {Shang}, \citenamefont {Zhang}, \citenamefont {Liu}, \citenamefont {Chen}, \citenamefont {Yang}, \citenamefont {Li}, \citenamefont {Li}, \citenamefont {Zhang}, \citenamefont {Xiong}, \citenamefont {Liu} \emph {et~al.}}]{shang2018surface}%
  \BibitemOpen
  \bibfield  {author} {\bibinfo {author} {\bibfnamefont {Q.}~\bibnamefont {Shang}}, \bibinfo {author} {\bibfnamefont {S.}~\bibnamefont {Zhang}}, \bibinfo {author} {\bibfnamefont {Z.}~\bibnamefont {Liu}}, \bibinfo {author} {\bibfnamefont {J.}~\bibnamefont {Chen}}, \bibinfo {author} {\bibfnamefont {P.}~\bibnamefont {Yang}}, \bibinfo {author} {\bibfnamefont {C.}~\bibnamefont {Li}}, \bibinfo {author} {\bibfnamefont {W.}~\bibnamefont {Li}}, \bibinfo {author} {\bibfnamefont {Y.}~\bibnamefont {Zhang}}, \bibinfo {author} {\bibfnamefont {Q.}~\bibnamefont {Xiong}}, \bibinfo {author} {\bibfnamefont {X.}~\bibnamefont {Liu}}, \emph {et~al.},\ }\bibfield  {title} {\bibinfo {title} {Surface plasmon enhanced strong exciton--photon coupling in hybrid inorganic--organic perovskite nanowires},\ }\href@noop {} {\bibfield  {journal} {\bibinfo  {journal} {Nano letters}\ }\textbf {\bibinfo {volume} {18}},\ \bibinfo {pages} {3335} (\bibinfo {year} {2018})}\BibitemShut {NoStop}%
\bibitem [{\citenamefont {Canales}\ \emph {et~al.}(2021)\citenamefont {Canales}, \citenamefont {Baranov}, \citenamefont {Antosiewicz},\ and\ \citenamefont {Shegai}}]{Canales2021}%
  \BibitemOpen
  \bibfield  {author} {\bibinfo {author} {\bibfnamefont {A.}~\bibnamefont {Canales}}, \bibinfo {author} {\bibfnamefont {D.~G.}\ \bibnamefont {Baranov}}, \bibinfo {author} {\bibfnamefont {T.~J.}\ \bibnamefont {Antosiewicz}},\ and\ \bibinfo {author} {\bibfnamefont {T.}~\bibnamefont {Shegai}},\ }\bibfield  {title} {\bibinfo {title} {{Abundance of cavity-free polaritonic states in resonant materials and nanostructures}},\ }\href@noop {} {\bibfield  {journal} {\bibinfo  {journal} {J. Chem. Phys.}\ }\textbf {\bibinfo {volume} {154}},\ \bibinfo {pages} {024701} (\bibinfo {year} {2021})}\BibitemShut {NoStop}%
\bibitem [{\citenamefont {Weber}\ \emph {et~al.}(2023)\citenamefont {Weber}, \citenamefont {K{\"u}hner}, \citenamefont {Sortino}, \citenamefont {Ben~Mhenni}, \citenamefont {Wilson}, \citenamefont {K{\"u}hne}, \citenamefont {Finley}, \citenamefont {Maier},\ and\ \citenamefont {Tittl}}]{Weber2023}%
  \BibitemOpen
  \bibfield  {author} {\bibinfo {author} {\bibfnamefont {T.}~\bibnamefont {Weber}}, \bibinfo {author} {\bibfnamefont {L.}~\bibnamefont {K{\"u}hner}}, \bibinfo {author} {\bibfnamefont {L.}~\bibnamefont {Sortino}}, \bibinfo {author} {\bibfnamefont {A.}~\bibnamefont {Ben~Mhenni}}, \bibinfo {author} {\bibfnamefont {N.~P.}\ \bibnamefont {Wilson}}, \bibinfo {author} {\bibfnamefont {J.}~\bibnamefont {K{\"u}hne}}, \bibinfo {author} {\bibfnamefont {J.~J.}\ \bibnamefont {Finley}}, \bibinfo {author} {\bibfnamefont {S.~A.}\ \bibnamefont {Maier}},\ and\ \bibinfo {author} {\bibfnamefont {A.}~\bibnamefont {Tittl}},\ }\bibfield  {title} {\bibinfo {title} {Intrinsic strong light-matter coupling with self-hybridized bound states in the continuum in van der waals metasurfaces},\ }\href@noop {} {\bibfield  {journal} {\bibinfo  {journal} {Nature Materials}\ }\textbf {\bibinfo {volume} {22}},\ \bibinfo {pages} {970} (\bibinfo {year} {2023})}\BibitemShut {NoStop}%
\bibitem [{\citenamefont {Blanco}\ and\ \citenamefont {{Garc{\'{i}}a De Abajo}}(2004)}]{Blanco2004}%
  \BibitemOpen
  \bibfield  {author} {\bibinfo {author} {\bibfnamefont {L.~A.}\ \bibnamefont {Blanco}}\ and\ \bibinfo {author} {\bibfnamefont {F.~J.}\ \bibnamefont {{Garc{\'{i}}a De Abajo}}},\ }\bibfield  {title} {\bibinfo {title} {{Spontaneous light emission in complex nanostructures}},\ }\href@noop {} {\bibfield  {journal} {\bibinfo  {journal} {Physical Review B - Condensed Matter and Materials Physics}\ }\textbf {\bibinfo {volume} {69}},\ \bibinfo {pages} {205414} (\bibinfo {year} {2004})}\BibitemShut {NoStop}%
\bibitem [{\citenamefont {Anger}\ \emph {et~al.}(2006)\citenamefont {Anger}, \citenamefont {Bharadwaj},\ and\ \citenamefont {Novotny}}]{anger2006enhancement}%
  \BibitemOpen
  \bibfield  {author} {\bibinfo {author} {\bibfnamefont {P.}~\bibnamefont {Anger}}, \bibinfo {author} {\bibfnamefont {P.}~\bibnamefont {Bharadwaj}},\ and\ \bibinfo {author} {\bibfnamefont {L.}~\bibnamefont {Novotny}},\ }\bibfield  {title} {\bibinfo {title} {Enhancement and quenching of single-molecule fluorescence},\ }\href@noop {} {\bibfield  {journal} {\bibinfo  {journal} {Physical review letters}\ }\textbf {\bibinfo {volume} {96}},\ \bibinfo {pages} {113002} (\bibinfo {year} {2006})}\BibitemShut {NoStop}%
\bibitem [{\citenamefont {Pelton}(2015)}]{Pelton2015}%
  \BibitemOpen
  \bibfield  {author} {\bibinfo {author} {\bibfnamefont {M.}~\bibnamefont {Pelton}},\ }\bibfield  {title} {\bibinfo {title} {{Modified spontaneous emission in nanophotonic structures}},\ }\href@noop {} {\bibfield  {journal} {\bibinfo  {journal} {Nature Photonics}\ }\textbf {\bibinfo {volume} {9}},\ \bibinfo {pages} {427} (\bibinfo {year} {2015})}\BibitemShut {NoStop}%
\bibitem [{\citenamefont {Chikkaraddy}\ \emph {et~al.}(2016)\citenamefont {Chikkaraddy}, \citenamefont {{De Nijs}}, \citenamefont {Benz}, \citenamefont {Barrow}, \citenamefont {Scherman}, \citenamefont {Rosta}, \citenamefont {Demetriadou}, \citenamefont {Fox}, \citenamefont {Hess},\ and\ \citenamefont {Baumberg}}]{Chikkaraddy2016}%
  \BibitemOpen
  \bibfield  {author} {\bibinfo {author} {\bibfnamefont {R.}~\bibnamefont {Chikkaraddy}}, \bibinfo {author} {\bibfnamefont {B.}~\bibnamefont {{De Nijs}}}, \bibinfo {author} {\bibfnamefont {F.}~\bibnamefont {Benz}}, \bibinfo {author} {\bibfnamefont {S.~J.}\ \bibnamefont {Barrow}}, \bibinfo {author} {\bibfnamefont {O.~A.}\ \bibnamefont {Scherman}}, \bibinfo {author} {\bibfnamefont {E.}~\bibnamefont {Rosta}}, \bibinfo {author} {\bibfnamefont {A.}~\bibnamefont {Demetriadou}}, \bibinfo {author} {\bibfnamefont {P.}~\bibnamefont {Fox}}, \bibinfo {author} {\bibfnamefont {O.}~\bibnamefont {Hess}},\ and\ \bibinfo {author} {\bibfnamefont {J.~J.}\ \bibnamefont {Baumberg}},\ }\bibfield  {title} {\bibinfo {title} {{Single-molecule strong coupling at room temperature in plasmonic nanocavities}},\ }\href@noop {} {\bibfield  {journal} {\bibinfo  {journal} {Nature}\ }\textbf {\bibinfo {volume} {535}},\ \bibinfo {pages} {127} (\bibinfo {year} {2016})}\BibitemShut {NoStop}%
\bibitem [{\citenamefont {Park}\ \emph {et~al.}(2019)\citenamefont {Park}, \citenamefont {May}, \citenamefont {Leng}, \citenamefont {Wang}, \citenamefont {Kropp}, \citenamefont {Gougousi}, \citenamefont {Pelton},\ and\ \citenamefont {Raschke}}]{Park2019}%
  \BibitemOpen
  \bibfield  {author} {\bibinfo {author} {\bibfnamefont {K.~D.}\ \bibnamefont {Park}}, \bibinfo {author} {\bibfnamefont {M.~A.}\ \bibnamefont {May}}, \bibinfo {author} {\bibfnamefont {H.}~\bibnamefont {Leng}}, \bibinfo {author} {\bibfnamefont {J.}~\bibnamefont {Wang}}, \bibinfo {author} {\bibfnamefont {J.~A.}\ \bibnamefont {Kropp}}, \bibinfo {author} {\bibfnamefont {T.}~\bibnamefont {Gougousi}}, \bibinfo {author} {\bibfnamefont {M.}~\bibnamefont {Pelton}},\ and\ \bibinfo {author} {\bibfnamefont {M.~B.}\ \bibnamefont {Raschke}},\ }\bibfield  {title} {\bibinfo {title} {{Tip-enhanced strong coupling spectroscopy, imaging, and control of a single quantum emitter}},\ }\href@noop {} {\bibfield  {journal} {\bibinfo  {journal} {Science Advances}\ }\textbf {\bibinfo {volume} {5}},\ \bibinfo {pages} {eaav5931} (\bibinfo {year} {2019})}\BibitemShut {NoStop}%
\bibitem [{\citenamefont {Hu}\ \emph {et~al.}(2021)\citenamefont {Hu}, \citenamefont {Shi}, \citenamefont {Zhang},\ and\ \citenamefont {Xu}}]{Hu2021}%
  \BibitemOpen
  \bibfield  {author} {\bibinfo {author} {\bibfnamefont {H.}~\bibnamefont {Hu}}, \bibinfo {author} {\bibfnamefont {Z.}~\bibnamefont {Shi}}, \bibinfo {author} {\bibfnamefont {S.}~\bibnamefont {Zhang}},\ and\ \bibinfo {author} {\bibfnamefont {H.}~\bibnamefont {Xu}},\ }\bibfield  {title} {\bibinfo {title} {{Unified treatment of scattering, absorption, and luminescence spectra from a plasmon-exciton hybrid by temporal coupled-mode theory}},\ }\href@noop {} {\bibfield  {journal} {\bibinfo  {journal} {Journal of Chemical Physics}\ }\textbf {\bibinfo {volume} {155}},\ \bibinfo {pages} {074104} (\bibinfo {year} {2021})}\BibitemShut {NoStop}%
\bibitem [{\citenamefont {Ringler}\ \emph {et~al.}(2008)\citenamefont {Ringler}, \citenamefont {Schwemer}, \citenamefont {Wunderlich}, \citenamefont {Nichtl}, \citenamefont {K{\"u}rzinger}, \citenamefont {Klar},\ and\ \citenamefont {Feldmann}}]{ringler2008shaping}%
  \BibitemOpen
  \bibfield  {author} {\bibinfo {author} {\bibfnamefont {M.}~\bibnamefont {Ringler}}, \bibinfo {author} {\bibfnamefont {A.}~\bibnamefont {Schwemer}}, \bibinfo {author} {\bibfnamefont {M.}~\bibnamefont {Wunderlich}}, \bibinfo {author} {\bibfnamefont {A.}~\bibnamefont {Nichtl}}, \bibinfo {author} {\bibfnamefont {K.}~\bibnamefont {K{\"u}rzinger}}, \bibinfo {author} {\bibfnamefont {T.}~\bibnamefont {Klar}},\ and\ \bibinfo {author} {\bibfnamefont {J.}~\bibnamefont {Feldmann}},\ }\bibfield  {title} {\bibinfo {title} {Shaping emission spectra of fluorescent molecules with single plasmonic nanoresonators},\ }\href@noop {} {\bibfield  {journal} {\bibinfo  {journal} {Physical review letters}\ }\textbf {\bibinfo {volume} {100}},\ \bibinfo {pages} {203002} (\bibinfo {year} {2008})}\BibitemShut {NoStop}%
\bibitem [{\citenamefont {Liao}\ \emph {et~al.}(2019)\citenamefont {Liao}, \citenamefont {Shan},\ and\ \citenamefont {Li}}]{liao2019situ}%
  \BibitemOpen
  \bibfield  {author} {\bibinfo {author} {\bibfnamefont {M.}~\bibnamefont {Liao}}, \bibinfo {author} {\bibfnamefont {B.}~\bibnamefont {Shan}},\ and\ \bibinfo {author} {\bibfnamefont {M.}~\bibnamefont {Li}},\ }\bibfield  {title} {\bibinfo {title} {In situ raman spectroscopic studies of thermal stability of all-inorganic cesium lead halide (cspbx3, x= cl, br, i) perovskite nanocrystals},\ }\href@noop {} {\bibfield  {journal} {\bibinfo  {journal} {The journal of physical chemistry letters}\ }\textbf {\bibinfo {volume} {10}},\ \bibinfo {pages} {1217} (\bibinfo {year} {2019})}\BibitemShut {NoStop}%
\bibitem [{\citenamefont {Garcia-Vergara}\ \emph {et~al.}(2017)\citenamefont {Garcia-Vergara}, \citenamefont {Dem{\'e}sy},\ and\ \citenamefont {Zolla}}]{garcia2017extracting}%
  \BibitemOpen
  \bibfield  {author} {\bibinfo {author} {\bibfnamefont {M.}~\bibnamefont {Garcia-Vergara}}, \bibinfo {author} {\bibfnamefont {G.}~\bibnamefont {Dem{\'e}sy}},\ and\ \bibinfo {author} {\bibfnamefont {F.}~\bibnamefont {Zolla}},\ }\bibfield  {title} {\bibinfo {title} {Extracting an accurate model for permittivity from experimental data: hunting complex poles from the real line},\ }\href@noop {} {\bibfield  {journal} {\bibinfo  {journal} {Optics Letters}\ }\textbf {\bibinfo {volume} {42}},\ \bibinfo {pages} {1145} (\bibinfo {year} {2017})}\BibitemShut {NoStop}%
\bibitem [{\citenamefont {Wu}\ \emph {et~al.}(2023)\citenamefont {Wu}, \citenamefont {Arrivault}, \citenamefont {Yan},\ and\ \citenamefont {Lalanne}}]{wu2023modal}%
  \BibitemOpen
  \bibfield  {author} {\bibinfo {author} {\bibfnamefont {T.}~\bibnamefont {Wu}}, \bibinfo {author} {\bibfnamefont {D.}~\bibnamefont {Arrivault}}, \bibinfo {author} {\bibfnamefont {W.}~\bibnamefont {Yan}},\ and\ \bibinfo {author} {\bibfnamefont {P.}~\bibnamefont {Lalanne}},\ }\bibfield  {title} {\bibinfo {title} {Modal analysis of electromagnetic resonators: user guide for the man program},\ }\href@noop {} {\bibfield  {journal} {\bibinfo  {journal} {Computer Physics Communications}\ }\textbf {\bibinfo {volume} {284}},\ \bibinfo {pages} {108627} (\bibinfo {year} {2023})}\BibitemShut {NoStop}%
\bibitem [{\citenamefont {Sauvan}\ \emph {et~al.}(2022)\citenamefont {Sauvan}, \citenamefont {Wu}, \citenamefont {Zarouf}, \citenamefont {Muljarov},\ and\ \citenamefont {Lalanne}}]{sauvan2022normalization}%
  \BibitemOpen
  \bibfield  {author} {\bibinfo {author} {\bibfnamefont {C.}~\bibnamefont {Sauvan}}, \bibinfo {author} {\bibfnamefont {T.}~\bibnamefont {Wu}}, \bibinfo {author} {\bibfnamefont {R.}~\bibnamefont {Zarouf}}, \bibinfo {author} {\bibfnamefont {E.~A.}\ \bibnamefont {Muljarov}},\ and\ \bibinfo {author} {\bibfnamefont {P.}~\bibnamefont {Lalanne}},\ }\bibfield  {title} {\bibinfo {title} {Normalization, orthogonality, and completeness of quasinormal modes of open systems: the case of electromagnetism},\ }\href@noop {} {\bibfield  {journal} {\bibinfo  {journal} {Optics Express}\ }\textbf {\bibinfo {volume} {30}},\ \bibinfo {pages} {6846} (\bibinfo {year} {2022})}\BibitemShut {NoStop}%
\bibitem [{\citenamefont {Kavokin}\ \emph {et~al.}(2017)\citenamefont {Kavokin}, \citenamefont {Baumberg}, \citenamefont {Malpuech},\ and\ \citenamefont {Laussy}}]{kavokin2017microcavities}%
  \BibitemOpen
  \bibfield  {author} {\bibinfo {author} {\bibfnamefont {A.}~\bibnamefont {Kavokin}}, \bibinfo {author} {\bibfnamefont {J.~J.}\ \bibnamefont {Baumberg}}, \bibinfo {author} {\bibfnamefont {G.}~\bibnamefont {Malpuech}},\ and\ \bibinfo {author} {\bibfnamefont {F.~P.}\ \bibnamefont {Laussy}},\ }\href@noop {} {\emph {\bibinfo {title} {Microcavities}}}\ (\bibinfo  {publisher} {Oxford University Press},\ \bibinfo {address} {198 Madison Avenue, New York, NY 10016, United States of America},\ \bibinfo {year} {2017})\BibitemShut {NoStop}%
\bibitem [{\citenamefont {Filip}\ \emph {et~al.}(2021)\citenamefont {Filip}, \citenamefont {Haber},\ and\ \citenamefont {Neaton}}]{filip2021phonon}%
  \BibitemOpen
  \bibfield  {author} {\bibinfo {author} {\bibfnamefont {M.~R.}\ \bibnamefont {Filip}}, \bibinfo {author} {\bibfnamefont {J.~B.}\ \bibnamefont {Haber}},\ and\ \bibinfo {author} {\bibfnamefont {J.~B.}\ \bibnamefont {Neaton}},\ }\bibfield  {title} {\bibinfo {title} {Phonon screening of excitons in semiconductors: Halide perovskites and beyond},\ }\href@noop {} {\bibfield  {journal} {\bibinfo  {journal} {Physical review letters}\ }\textbf {\bibinfo {volume} {127}},\ \bibinfo {pages} {067401} (\bibinfo {year} {2021})}\BibitemShut {NoStop}%
\end{thebibliography}%

\end{document}